\documentclass{report}
\usepackage[xdvi]{epsfig}
\begin {document}
\title {An Equilibrium Balance of the Universe}
\author{Ernst Fischer}
\date{e-mail: e.fischer.stolberg@t-online.de}
\maketitle
\tableofcontents
\pagestyle{plain}
\chapter{Introduction}
During the last decades our knowledge of the structure of the universe has
taken an enormous step forward due to improved methods of observation and the
development of new devices. At the same time the theoretical models to
describe the structure and history of the universe have been developed
further. Today there exists a so called 'concordance model' that can describe
the development of the cosmos, beginning from a 'big bang' up to the presence
and even forecasting its future.

But in spite of the ability of the mathematical model, to describe many of
the observed phenomena rather accurately, some people feel uncomfortable with
this model, as it requires rigourous changes of our accustomed thinking.
Already Einstein's idea, to abandon space and time as an absolute reference
system, was difficult to accept. But what is still worse with the concordance
model is, that it shifts the conditions, which led to the present day state
of the universe, into some initial conditions and physical principles, which
can neither be proved nor disproved observationally, as they only prevailed
at some instant of time, when our known physical laws were not valid. In
addition the postulated existence of matter or energy fields with rather
strange properties, which should dominate the development of the universe,
but which cannot be observed in earthbound experiments, makes the model
questionable.

But it is the initial singularity, the 'big bang', which appears scarcely
acceptable from the physical point of view. Of course, mathematics can easily
handle singularities, and in some cases it may be convenient to use singular
distributions to describe some effect mathematically, but physically a state
of infinite density of some field appears impossible. In physics all our
experience says that every model theory, which leads to singularities, is
wrong or at least inaccurate under the corresponding conditions. With the big
bang theory also the excuse does not help much that near the singularity the
known laws of physics are no longer valid. As long as we know no alternative
laws, the value of such a statement is zero. Besides, singularities are not
restricted to the initial condition, but they are the inevitable final state
of matter, when a star or galaxy collapses into a 'black hole' under the
influence of gravitation.

According to the generally discussed model of today the size of the universe
is infinite, but at the same time it is expanding as a whole. While in the
static model, which was originally proposed by Einstein, the curvature of
space could explain in an elegant way, why the universe can be of finite
size, but at the same time has no limits, this is not the case with the
'concordance model'.

Besides of that, to remain in agreement with observations, the 'concordance
model' requires the existence of completely new ingredients to the universe
with properties, which are different from all, what was known by now. 95\% of
gravitational interaction should be dominated by dark matter and dark energy,
some stuff, which interacts with conventional matter only by gravity. But the
formation of this stuff in the big bang remains a riddle, just as the force
or interaction which finally drives the expansion of the universe.

But nevertheless with a suitable adjustment of parameters the 'concordance
model' has succeeded to create a system, which is able to describe a large
number of observations. As well the formation and abundance of chemical
elements as the cosmological microwave background and its inhomogeneities,
the red shift of spectral lines and the formation of observed structures can
be explained. But it is just this explanation of cosmic structures, which
runs into difficulties with increasing improvement of observations. It is
scarcely possible to explain, how shortly after the big bang galaxies and
galaxy clusters should have been formed, which have similar appearance and
metal abundance as those of today.

The concordance model in its present form gives an acceptable description of
the observational material, but on the expense of many assumptions, which
cannot be verified or falsified and which are scarcely acceptable by our
accustomed physical understanding. We must put the question, if there is no
alternative to this model, which does not need so many unprovable assumptions
and singularities. Hasn't an error crept in already into the basic
assumptions, which finally leads us into a blind alley? Aren't we in a state
like Ptolemaeus, who tried to keep his geocentric universe in accordance with
observations by adding more and more epicycles?

In this paper we will try to find possible causes for such a erroneous
development and to propose a concept, by which an alternative to the big bang
model is possible, which avoids singularity problems and does not need the
existence of the mysterious dark matter or dark energy, though explaining all
the observations. Of course, it is scarcely possible within the scope of this
paper to examine all observations, which have been made over many years, with
respect to their compatibility with the model presented here. Instead we will
propose a new way of thinking and a new interpretation of observed facts,
which may finally lead to an improved understanding of the universe.

We will not concentrate on detailed observations and dispense most references
to literature. Literature in the field of astrophysics and cosmology is so
copious that a nearly complete reference list would be out of the scope of
this paper. Only where we directly refer to data or results from individual
authors, references are given.

\chapter{Back to Einstein}
The probably most important step towards our present day cosmological models
was the development of the General Theory of Relativity (GRT) by Albert
Einstein. His starting point, to use the equivalence of gravitational mass
and inertial mass to describe the complete gravitational interaction by the
geometrical properties of the space-time continuum, led to completely new
understanding of space and time.

Already with the Special Theory of Relativity and its various experimental
confirmations it had become clear that space and time coordinates do not have
an absolute meaning, but only the relation of bodies with respect to each
other. With his General Theory of Relativity Einstein succeeded in describing
the dynamics matter and energy under the influence of gravitation in a way,
independent of the reference system, which is only determined by the metric
of the space-time continuum.

By the methods of GRT it was not only possible to explain some phenomena like
the shift of perihelion of Mercury or the aberration of starlight passing the
sun, the new theory offered also the opportunity to make statements on the
universe as a whole. With the assumption that space is curved it appeared
possible that the universe has no limit, but in spite of that its size is
finite. The view of an infinite size of space, which appeared uncomfortable
for most physicist of that time, disappeared with Einstein's model. Of
course, this model demanded the introduction of a new fundamental quantity,
the so called 'cosmological constant', which led to a negative pressure and
thus to a repulsive force, counteracting the attractive gravitational force
of matter.

A few years after publication of Einstein's theory the mathematician
Friedmann showed that there existed a time-dependent solution of Einstein's
field equations without the necessity of a cosmological constant under the
proposition that space is continuously expanding. As just at the same time
the red shift of spectral lines of distant galaxies was discovered, it
appeared obvious to take this as a prove that the universe is continuously
expanding. Also Einstein himself changed his mind to this interpretation,
especially as his original static model proved unstable within the framework
of time-dependent solutions of the basic equations. With a cosmological
constant it was possible to establish an equilibrium to the gravitational
attraction at a certain radius of curvature, but any small deviation of the
curvature would lead to an irreversible expansion or to collapse. Later
Einstein defined the introduction of the cosmological constant as his
'greatest blunder'.

With the acceptance of the paradigm of the expanding universe practically the
path to the modern 'concordance model' was traced out. The starting point as
a singularity and the limited time, available for the development of
structures from the big bang to the presence, was fixed. All the further
corrections, as for instance the addition of an inflationary phase of
expansion, or in the last years the claim for the existence of dark energy,
were only necessary to keep the model in accordance with the enhanced
accuracy of observations.

Inflation, the exponential expansion in the beginning, was necessary to
explain, why all parts of the universe have developed in a similar way. That
means that they must have had a causal connection, though this was impossible
in later times due to the expansion of space. But on the other side
introduction of the inflation model required that the total density of all
matter, which contributes to gravitation, should equal a critical density,
which, however, cannot be explained by the observed matter nor by the
chemical element abundances resulting from the big bang nucleosynthesis.

To save the matter balance then the existence of some 'dark matter' was
claimed, which should contribute to gravitation, but not interact with normal
matter or radiation in any other way. But even this correction was not
sufficient, as such matter would cluster in the same way as visible matter
due to gravitational interaction. But as from new observations the existence
of this dark matter in quantities, necessary to reach the critical density of
the universe, must be excluded, finally 'dark energy' was created, some stuff
which brings the matter balance to the critical value, but does not
conglomerate, as it possesses a negative pressure and thus a repulsive
gravitation. With this new form of matter we are practically back with
Einstein's cosmological constant without having a better explanation than
Einstein had ninety years ago.

If we want to find out the origin of any failure in the development, we must
go back further than to the starting point of the theory of an expanding
universe. The basic assumptions of GRT should be scrutinised. Of course,
since the time of Einstein until today this theory has been confirmed with
steadily increasing accuracy, but all these confirmations are related to
parameter regions, where small deviations from Newtonian theory of
gravitation are discussed. The basics, which were confirmed, were changes of
distances or running times of signals, caused by the curvature of space-time,
and the existence of gravitational radiation resulting from the fact that
changes of the metric of space-time are limited by the speed of light.

There is however one conceptual problem in GRT, to which we should direct our
attention: the definition of density. In pre-relativistic times we were
accustomed to define the density of a quantity by the ratio of the amount of
this quantity and of the volume element in which it is contained. This ratio
or, for spatially varying quantities its limiting value, when the size of the
volume element approaches zero, could be considered as the local density
value of the corresponding quantity.

According to the concept of GRT, however, the size of a volume element
depends on the metric and thus on the distribution of matter or energy in the
surrounding space. Thus there is no longer a unique definition of density in
the conventional sense. But Einstein easily found a way out of this
situation. As a local quantity, which should enter the field equations, he
defined the density in the 'tangential space', that means the density value
one would measure, if all the surrounding masses were removed to infinity.

In this way the problem of a unique definition is solved, but at the cost of
another one. Calculation of the integral of the so defined density over the
region of an extended mass distribution, the result is different from what we
would get by counting the number of atoms and multiplying it with their
characteristic mass. For the example of a spherically symmetric mass
distribution, for which the exact solution of the Einstein equations is
known, to the first approximation the difference is just the value of
potential energy, that means the binding energy, which according to Newtonian
theory has to be supplied, to distribute the mass into infinite space against
the action of gravitation (see appendix for mathematical details).

But which of the two different values of the mass determines the
gravitational action to the outside? Is it, as Einstein said, that 'the
inertia of mass is enhanced, when ponderable matter is accumulated in its
surrounding' (and according to the principle of equivalence also the
gravitational mass), remains it unaffected or does it eventually decrease?

The easiest way to answer this question from our experience is, to consider
the limiting case of the collapse of a matter distribution under the
influence of gravitation, which finally ends up in an extremely compact
state, according to present day concepts, even in a singularity. If the
gravitational interaction with the outside world would increase due to the
enhanced binding energy, the dynamics of the universe would be governed
completely by such singular systems, a fact, which contradicts all
observations. But a decreasing gravitational interaction appears improbable
as well. In this case is should approach zero in a theoretical model which
contains no characteristic scale length. Observations show, however, that
there exist very compact objects in space with gravitational masses,
corresponding to those of stars or even millions of stars.

Thus we must conclude that gravitational interaction of matter to the outside
remains unchanged, when the internal structure is changed. It cannot be true
that it is the density, defined in the Euclidean tangent space and integrated
over the volume of the curved space, which determines the gravitation to the
outside. We have to subtract the binding energy from the matter density to
obtain the correct gravitating mass after integration.

The energy tensor, which was used by Einstein to describe the universe as a
whole, does not contain the binding energy or, to express it more generally,
the (negative) potential energy in any form. Instead he takes over the energy
tensor for homogeneously distributed matter from special theory of
relativity, in which only the density of matter and the kinetic energy of the
particles, that means the pressure, contribute to the total energy. Of
course, this can be regarded as an allowed approximation, as long as the
metric and, resulting from this, the dynamics of test particles is dominated
by the matter distribution. For the aberration of light, passing the sun, the
contribution of potential energy compared to the mass-energy is less then
$10^{-6}$ of the total action at maximum. Of similar size are the corrections
also with other observations which have been used as confirmations of the
general theory of relativity.

But neglecting potential energy in the energy tensor, which appears as the
tiny correction in this case, has strong consequences for the general
concept. According to the general theory of relativity, as it is
conventionally used, conservation laws of energy and momentum are valid only
locally, while changes of the metric can generate or destroy energy. The fact
that the dynamical behaviour of test particles within the gravitational field
of surrounding matter is described correctly is no indication, however, that
the energy tensor, the source term of the field equations, is fully correct.
Only by adding potential energy it is possible to develop a model, in which,
as in Newtonian theory, the conservation laws are valid locally and in
integral form as well.

But the question is: how can we include potential energy within the scope of
a covariant theory, that means, in a form which is independent from the
reference system? To fulfil this condition any formula for the potential
energy may contain only locally defined variables of state beside of the
metric tensor and quantities derived from it without differentiation. In
addition potential energy should vanish in empty space, that means, it should
be proportional to the density of matter. It should vanish in the limiting
case of Euclidean geometry, where the special theory of relativity is valid,
and in the limiting case of weak curvature potential energy must result in
analogy to Newtonian theory (for the mathematical formulation see appendix).

Formally the only sensible possibility to fulfil these conditions is an
expression, which for a spatially homogeneous solution looks like the
'cosmological term' in Einstein's static universe, that means, a product of a
scalar quantity and the metric tensor. But now the scalar factor is no longer
a constant, but it is a quantity, which is proportional to the local energy
or matter density and to the inverse of the radius of curvature. The
essential difference to the 'cosmological constant' is the fact that an
expansion of the universe, that means, a change of curvature, affects
potential energy more than mass density. As a consequence a static universe
is stable, contrary to the unstable Einstein universe, where the
'cosmological constant' can balance gravitational attraction only at one
fixed value of curvature.

There is, however on more serious difference. The field equations of general
relativity are relations between tensors. Every tensor component has to
fulfil the corresponding equality conditions. But the assumption of a
homogeneous medium reduces the number of degrees of freedom to such an extent
that it poses additional conditions to the possible forms of the energy
tensor, which must be satisfied to allow time dependent solutions. The energy
fields have to obey certain equations of state. Only a pressureless mass
distribution can change proportional to the volume. Radiation fields, that
means, fields with pressure proportional to 1/3 of the energy density, must
vary with the fourth power of the scale variable. The cosmological term with
its negative pressure must be constant, independent of any expansion.

The expression, which we follows from our considerations as a reasonable
description of potential energy, does not fulfil this condition. With other
words: all time dependent solutions of the field equations for a homogeneous
universe lead to contradiction. There is no solution but the trivial one, in
which the potential energy of matter is in equilibrium with matter itself.
The decisive fact is that potential energy due to its negative pressure acts
as a repulsive force counteracting gravitating attraction of matter.

The conception of negative pressure is not as strange as one would think.
Also in Newtons theory of gravitation the attraction of masses generates a
negative pressure. What is different in general relativity is the fact that
this pressure contributes to the gravitational force.

We can imagine the action of negative pressure in Newtonian physics from a
simple thought experiment. If we think of space, homogeneously filled with
particles at rest, every particle is in equilibrium with its neighbouring
masses, as the gravitational forces compensate each other. If, however, the
homogeneous distribution contains a little cavity which does not contain
matter, a 'vacuum bubble', equilibrium is distorted for the particles near
the boundary of the bubble. There are less particles pulling inwards than
outwards, so that there is a resulting force tending to enhance the volume of
the bubble, as if the bubble had a positive pressure compared to the mass
filled space. If we assign zero pressure to the vacuum, the pressure in the
surrounding medium must be considered negative.

That we never observe this negative pressure in practice is caused by the
fact that under all terrestrial conditions the positive kinetic pressure
outbalances it by many orders of magnitude. Thus any 'vacuum bubble' would be
immediately filled up by the motion of particles. Generally speaking,
positive pressure leads to a homogeneous distribution of matter, while
negative pressure leads to clumping and formation of structures. We shall
come back to this point later.

An essential difference to the 'big bang' model should be mentioned here. In
its presently preferred form this model contains a term, which formally acts
like Einstein's cosmological constant. Most people ascribe this term to the
zero point energy of quantum fields. But there exists no reasonable answer to
the question, why this term should be different from zero, but smaller by
many orders of magnitude than estimated by any quantum mechanical model.

In the equilibrium model considered here this problem does not exist. Any
energy term, which has the form of a cosmological constant, contributes
exactly equal amounts to the density of positive energy and to the negative
potential energy. Thus it is without any effect in the field equations (see
appendix).

As an important result of the considerations in this chapter we have to state
that a universe, in which conservation of mass and energy is valid not only
locally, but also globally, has to be generally static, if it is homogeneous
except for local fluctuations. Thus there cannot be a 'big bang' and all the
structures which we observe must be regarded as local distortions of
homogeneity, as 'cosmic weather'. If this view of the universe shall be
regarded as correct, it must also give explanations for all the observations,
which have been explained by the 'concordance model' more or less
successfully. What is the reason for the cosmological red shift, if not
expansion of space? How can we explain the abundance of elements? What is the
source of the microwave background radiation, why is it so homogeneous and
what causes the slight inhomogeneities, and how have the matter structures
been formed?

In the next chapters we will try to answer these questions within the scope
of known and proved physical laws.

\chapter{Red shift}
The strongest argument, which has led to the development of the 'big bang'
model, was the discovery of the cosmological red shift, which was interpreted
as change of the wavelength of photons due to the general temporal change of
all length scales in the universe. On the other side this concept is based on
the assumption that with the propagation of light in a static universe the
wavelength of photons should remain unchanged. But this is an assumption,
which is not contained in the basic equations of general relativity, but
postulated in analogy to the force-free motion in Euclidean space.

As in Einstein's theory the motion of particles or quanta should completely
determined by the geometry of space-time, it appeared reasonable to assume
that the motion of a test particle should be analogous to the force-free
motion in Euclidean space. That means that analogous to the motion with
constant energy along a straight line, motion under the influence of
gravitation should follow a 'geodesic line', the straightest line in curved
space-time. That the energy of motion should be constant along this line does
not follow from the field equations. The derivation of the geodesic equation
from the principle of minimal action does not require that the action
integral is zero, but only that it is minimal.

Does a force-free motion actually exist in curved space? Iso`t every motion
along a curved path an accelerated motion? In Newtonian physics we have
learned that the action of compulsive forces is zero, as they act
perpendicular to the direction of motion. But this is true only, if the
action is immediate. One characteristic difference between Newtonian gravity
and general relativity is the fact that gravitational interaction is limited
to the speed of light, similar to electromagnetic interaction. As a
consequence the compulsive force has also a component in the direction of
motion. There exists a retardation of the interaction, as we know it from
electromagnetic theory.

Anyway the principle of geodesic motion can be regarded only as an
approximation. It is based on the assumption that a test mass is moving
inside a metric, which is determined by the distribution of other matter
fields. But at the same time every test mass is part of the matter
distribution and thus contributes to the metric. When in a initially
completely homogeneous universe some mass moves from A to B, the universe is
no longer homogeneous. It causes a change of the metric which propagates into
space at the speed of light.

In most cases this propagating 'gravitational wave' is so weak that to
measure it appears scarcely possible. Only with extremely strong changes of
the matter distribution, as they occur with the explosion of stars, we can
hope to register such waves with very sophisticated detectors.

The only observation which has by now confirmed the existence of
gravitational waves, as they are predicted by general relativity, and the
loss of energy associated with them, is the decrease of the rotation
frequency of the 'double pulsars'. The properties of these systems of
collapsed double stars can be accurately measured by their emission of
quasi-periodic electromagnetic signals. The observed frequency shift is
exactly as expected from general relativity. In this case the energy loss by
gravitational waves corresponds to the quadrupole radiation in
electromagnetism. Dipole radiation, which dominates in electromagnetic
radiation, cannot occur as gravitational radiation, as there are no negative
masses, which could form a dipole with the positive ones.

But there should be a 'monopole radiation', the equivalent to electromagnetic
bremsstrahlung. Every accelerated electric charge radiates energy, be it
bremsstrahlung, which we observe from interactions between free electrons and
ions in plasma, or the cyclotron radiation, which is generated by the
transversal acceleration, which occurs when electrons are fixed to a curved
path by magnetic fields.

Today there exists no exact derivation of this gravitational bremsstrahlung
from the field equations of general relativity. But with a simple model it is
possible to understand the general properties of this radiation. To do this
we start from Newtonian gravity, but regard the fact that the action of
forces has to be retarded due to the finite speed of interaction, like it is
done to calculate bremsstrahlung in electromagnetic interaction. To avoid the
singularities, which occur in Newtonian theory by integration over an
infinite volume, we assume that there exists a homogeneous curvature of space
so that the volume remains finite. Correspondingly the gravitational force
between two points should be assumed to be directed along the geodesic
connecting them instead of a straight line. The strength of force should
decrease with the square of the distance as measured along this line.

From theoretical point of view this mixture of Newtonian gravity with general
relativity may appear somewhat inaccurate, but it gives us the opportunity to
understand the mechanism of gravitational bremsstrahlung in a simple way.

Let us first consider a test mass in the middle between two other masses. The
gravitational forces of the two outer masses exerted on the test mass are in
equilibrium. This holds as well for Newtonian gravity as in general
relativity as long as the test mass is at rest.The retardation of forces is
equal for both outer masses. But things are different, if the test mass is in
motion. According to Newtonian gravity the equilibrium conditions do not
change, as the forces act instantaneously. Due to finite speed of signals in
general relativity, however, we have to consider the running time of the
signal. That means to calculate the forces we have to use the distances at an
earlier point of time. At that time the distance in the direction of motion
was larger and thus the attractive force was weaker and against the direction
of motion the effect is reversed. Thus there exists a resulting force
opposite to the direction of motion, which gets stronger with increasing
velocity of the test mass (for the mathematical details see appendix).

This principle can easily applied also to a test mass in the gravitational
field of an homogeneous matter distribution, if we include the modifications
of Newtonian theory, mentioned above, to consider the curvature of space.
Integration of the contributions of all volume elements to the force on a
moving test mass leads to loss of momentum, which is the higher the stronger
the curvature of space (see appendix). Apparently the principle of momentum
loss is valid for every motion with respect to some matter background which
can be regarded at rest. That means, it is valid not only in the
non-relativistic limit, but also for relativistic particles or energy quanta,
if inertia is expressed by its connection to momentum or energy.

This shows that in the theory of general relativity there exists a simple
explanation of the cosmological red shift, which does not need any expansion
of the universe. The momentum of photons and correspondingly also their
energy decreases in proportion to their actual momentum resp. energy. That
means: the frequency decreases exponentially with running time or distance
from the source, the wavelength increases correspondingly. As a first
approximation this is a linear increase of wavelength with distance, just the
same dependence as given from the big bang model and which is found from
observations.

In our picture the red shift is a consequence of the general principle of
momentum loss, which affects every motion, and which will prove necessary for
understanding the concept of the dynamical equilibrium model of the universe
presented here. One might argue that the loss of momentum or kinetic energy
does not agree with the existence of global energy conservation, called for
in the beginning. But here again we have the fact that motion leads to
equalisation of all contrasts of the matter distribution and thus to
diminishing of the negative potential energy, so that the global balance
remains unchanged. In the same way as during gravitational collapse kinetic
energy is produced on cost of potential energy, motion with respect to the
background leads to a more homogeneous distribution of matter and thus to a
diminished average of the negative gravitational potential.

Within a model, which takes a homogeneous universe a the starting point, such
an effect can scarcely be described, of course. But the fact that the total
energy of the universe remains constant, when the radiation field changes,
becomes clear from the following consideration:

We regard the universe as an homogeneous matter filled space. The curvature
of pace is determined by the density of matter. If we add to this space an
additional radiation field or a field of homogeneously distributed kinetic
energy, without taking into account the potential energy the Einstein field
equations would lead to an enhanced curvature of space. But this increase of
curvature is equivalent to a stronger negative potential and thus to an
increase of the negative potential energy of matter. This change of the
potential energy of matter just compensates the virtual energy supply by the
radiation field.

Of course, such an homogeneous creation of kinetic energy is purely
hypothetical, but it illustrates, how the global energy balance remains
unchanged, if a kinetic energy or radiation field is changing. In the real
world such exchange between potential and kinetic energy takes place as
localised fluctuations. Only when we discuss the effects which, within the
global equilibrium, lead to local instabilities and thus are responsible for
the formation or disintegration of structures, we have to consider this
exchange of kinetic and potential energy more in detail.

For the moment it is sufficient to state that within general theory of
relativity there is a clear explanation of the cosmological red shift which
does not require any expansion of the universe. Though an exact theoretical
analysis of the effect was not given here, it should be clear that momentum
loss for every motion in curved space is a necessary result of Einstein's
theory of gravitation.

\chapter{Black holes - the source of matter}
In the last chapters we have shown that the universe as a whole must be
static, if we assume that it is homogeneous on large scale, that means that
no region is preferred to others by special properties of the energy tensor.
The variety of cosmic phenomena, which we observe, results from local
fluctuations or perturbations of homogeneity. The interplay of structure
formation by attraction of masses, the tendency of kinetic energy to
establish a homogeneous distribution, and finally the repulsive action of the
gravitational potential lead to a continuous change of the local equilibria.

In this context it should be mentioned that according to the general theory
of relativity it is not the energy or matter density alone, which is
responsible for the attractive or repulsive action, but what counts is the
invariant trace of the energy tensor. In the case of matter at rest this is
only the energy equivalent of mass, that means, according to Einstein's
renowned formula $E=mc^2$. For moving masses or for a massless radiation
field we have to add the momentum part to the energy density, that means, the
pressure of a gas or the radiation pressure, respectively. This amounts to
one third of the energy density per degree of freedom, so that such fields
count double in the balance, compared to matter at rest. Just the other way
things are for potential energy. In this case the pressure is negative, so
that also the trace of the energy tensor is negative. Thus the action is
repulsive, and here again with twice the value compared to the energy
density.

When some matter field collapses, for instance by the formation of a star or
a galaxy, the process starts with the generation of kinetic energy of inward
motion, which is then transformed into thermal energy by collisions and thus
into gas pressure. At the same time with increasing pressure the potential
energy of matter is changed, so that the balance, as defined by general
relativity, remains unchanged.

The balance of a homogeneous universe is defined by the fact that the energy
equivalent of matter is in equilibrium with its own potential energy. Thus
the total energy content is equivalent to zero, if we regard total energy as
the trace of the energy tensor in the sense of general relativity. Locally
this equilibrium is unstable, however. Any local perturbation generates
energy of motion, which remains local, while the change of the gravitational
potential, the change of the metric, expands into space with the speed of
light and remains scarcely detectable locally.

Only in the limiting case of extremely high matter concentrations, when the
density of potential energy is comparable to the density of matter energy,
the repulsive action of the potential can stop the attractive force of
matter. According to the prevailing concepts of gravity contraction of matter
fields would continue into a singularity. A 'black hole' would form, as there
exists no force which could halt the collapse. The kinetic energy, which
matter takes up from the gravitational potential, would even enforce
collapse. But if the repulsive action of the potential energy is taken into
account, instead of a singularity under ideal conditions the final state is a
stable configuration, in which potential energy and mass energy are in
equilibrium. There are no 'black holes' in the sense of conventional theory,
regions, in which mass is inevitably lost from the active parts of the
universe.

'Black holes' must not be regarded as the end of cosmic evolution, but they
are the source, from which matter is continuously regenerated in the
universe. To understand this, we must look somewhat closer to their
properties. Most important for the understanding of the global mass balance
of the universe are the 'supermassive black holes', which are found in the
centre of most galaxies. From the dynamical behaviour of the surrounding
stars these appear to be very compact aggregations with million or even
billion times the solar mass.

According to Newtonian theory of gravitation galaxies are stable
configurations, in which angular momentum of matter together with the
pressure of gas and radiation are in equilibrium with the gravitational
force. Only by radiative and dissipative losses this equilibrium may be
disturbed, so that finally the system collapses and more and more matter is
accreted onto the central matter concentration.

According to general relativity there exists an additional loss mechanism:
gravitational radiation. As has been mentioned in the last chapter, for every
motion with respect to background of the universe gravitational radiation
reduces momentum and, of course, also angular momentum. This leads to a
continuous reduction of the rotational energy, so that on the time scale of
the Hubble constant every rotational motion is changed from a circular to a
spiral path, and finally all matter is collected in the centre of motion.

We must expect, however, that in the final stage of accretion the time scale
becomes so short that angular momentum is conserved approximately, so that
the 'black hole', which forms at the centre, will rotate rapidly.
Spectroscopic observations of the accretion disks of spiral galaxies have
strongly confirmed this picture of a rotating 'black hole'. Even details like
the broadening of spectral lines can well be explained by the accretion
model. But a fact, which was not well understood from conventional models, is
the emission of intense, highly relativistic matter jets, which are emitted
in the direction of the axis of rotation of many galaxies and which can be
observed along distances of many Mpc.

From dynamics of the accretion disk this mechanism cannot be explained, if we
assume that the inflowing matter inevitably vanishes into a black hole. But
if, as we have considered here, the final state of collapse is a
configuration, in which rest energy and kinetic energy form an equilibrium
with potential energy, the formation of these jets can easily be explained,
and we can predict some of their properties.

In the idealised case of an exactly spherical collapse a stable final
equilibrium configuration would exist, but by the continuous inflow of matter
this equilibrium will be disturbed. Increase of the total amount of matter
leads to an overshot of the repulsive force: matter must be expelled again.
The matter flowing inwards consists not only of distributed gas but complete
stars may be sucked in, bringing with them their angular momentum and the
surrounding magnetic fields. Even if the angular momenta and magnetic fields
of individual stars will compensate each other by part, one must assume that
after accretion of millions of stars and after the reduction to a very small
volume the resulting black hole will rotate rapidly and will be surrounded by
a strong magnetic field.

Charged matter particles can escape from such a configuration only from the
polar region, as otherwise their motion is bound to the closed magnetic field
lines. This effect gives us a simple explanation for the emission of strongly
collimated matter jets.

From the mechanism of formation we can deduce some of the properties of these
jets. We know that the matter is expelled from a region, in which potential
energy is of the same order as the rest energy of particles. As the final
stage of accretion is nearly adiabatic, the kinetic energy must be of the
same order, too. The energy per particle is so high that in this region only
the elementary building blocks of matter can exist, tat means, protons,
electrons, neutrons and neutrinos in equilibrium.

Correspondingly matter in the jet will first be composed from these
particles, too. The jet starts as an equilibrium plasma with the mean energy
per particle, corresponding to about the rest energy of protons, that means,
an equilibrium temperature of about $10^{13} K$. During expansion first the
plasma components can react with each other, but due to expansion inside the
jet the plasma becomes more and more diluted, so that the equilibrium is
frozen out with a composition which has been achieved by then. The jet
expands into space with this composition, but keeping its high temperature.

The formation mechanism of the composition equilibrium is practically the
same as adopted in big bang models to explain the synthesis of light
elements. Also there one assumes that the primordial composition of light
elements is fixed by freezing out from a hot, but expanding equilibrium
plasma. But whereas in big bang models one tries to draw conclusions on the
actual state of the universe from the singular event of elementary synthesis
in the beginning, this is not possible as easily, if formation of light
elements takes place continuously. They are formed at different locations
from black holes with different histories, which may have different masses
and may vary with respect to angular momentum and magnetic field strength.

But there remains the fact that black holes and their jets constitute a sink
of matter that has been processed in stars, and at the same time they are the
source of matter of primordial composition. Thus the formation of galaxies
and stars and the chemical processes during their evolution are no one way
street, but there exists also an opposite effect, which closes the matter
cycle of the universe.

\chapter{Ingredients of the intergalactic medium}

In the last chapter we have shown that gravitational collapse of galaxies
does not lead to singularities and to the final vanishing of matter from the
balance of change in the universe, bat that collapse leads to restoration of
the initial state of matter, from which new cosmic structures can be formed.
There exists a continuous cycle of matter from the hot plasma in the jets of
collapsed systems through the formation of large structures, the
fragmentation into individual galaxies or clusters of galaxies, the formation
of stars with their complex chemical and physical dynamics, to the renewed
gravitational collapse of stars or of complete galaxies, from which matter is
recovered in its initial state.

Before we start discussing details of this cycle, first we will consider the
properties and composition of the intergalactic medium, the properties of the
substrate, from which new structures can emerge. It is not only the matter,
expelled from black holes, which is distributed homogeneously into space.
Also radiation is emitted by various processes and fills the universe. In
addition large amounts of dust are produced by the explosion of stars, which
by part possess so much momentum that they can leave the galaxy, in which
they have been produced, and which are distributed into free space.

If the universe regarded as static on large scale there must be an
equilibrium of production and annihilation, not only for the hot plasma, but
also for the radiation field and the dust component. We will try to estimate,
how this equilibrium will look like, and show up, how these estimations are
confirmed by observations.

As a first step we will consider only those processes which determine the
global properties of the intergalactic medium, leaving out all detailed
structures. Only those processes are regarded, which can take place in a
homogeneously distributed medium. As the discussion below will show, this is
justified by the fact that the overwhelming part of matter is in such a state
of nearly homogeneous distribution.

\section{The intergalactic plasma}
Let us first look at the hot plasma component. As was discussed in the last
chapter, matter of primordial composition is continuously introduced by the
matter jets of quasars (or 'black holes'), mainly hydrogen and helium nuclei
and electrons, all with mean particle energies, which are in the order of the
proton rest energy. Though these particles carry electrical charge, due to
the high energy their interaction with the existing plasma is so weak that
the mean free path is several times the radius of the universe. The only
effective energy loss mechanism at that time is by gravitational radiation.
Only during several Hubble times the energy is diminished so far that the
mean free path, which decreases with the square of energy, becomes so short
that a considerable energy exchange with the existing plasma takes over. This
process then leads to a fast thermalisation of the plasma.

Thus we expect that the intergalactic plasma consists of a thermal component
and an additional component of very high energy. The equilibrium temperature
of the thermal component results from the balance of the rates, by which
energy is introduced from the matter jets and the loss rate by gravitational
radiation. Also after thermalisation gravitational radiation remains the
dominant loss mechanism.

Only when, caused by local fluctuations of density or temperature, losses by
electromagnetic bremsstrahlung become important, formation of structures will
start, which finally leads to galaxies and stars. Questions of structure
formation will be considered more in detail in the next chapter.

What should be stated here is that the period from the formation of matter in
the hot jets of quasars to the onset of structures takes many Hubble times,
while according to conventional view the lifetime of galaxies, the time from
formation of a galaxy to collapse into a black hole, is in the order of one
Hubble time. Thus we must expect that the dominant fraction of matter in the
universe is contained in the intergalactic plasma and only a small part is in
form of galaxies.

The model developed by now was only based on the theoretical considerations
without regarding the question, if these ideas are confirmed by observational
facts. By principle the existence of a hot plasma can be proved by its
optical properties, that means, by emission, absorption, and scattering of
electromagnetic radiation. But these measurements are very difficult, because
there exist no characteristic structures in the plasma, so that it is
scarcely possible to separate the effects from backgrounds of the measuring
device.

The only extinction effect, which might be measured in a fully ionised
plasma, is Compton scattering from free electrons. By this effect light from
distant galaxies or quasars should appear weakened, but the effect is so
small that it cannot be used to confirm or discard our model. Even for the
most distant quasars, observed by now, the extinction would just be in the
order of $10\%$, if we assume that the total density of the universe is close
to that, derived from the model presented here.

A more promising method is, to detect the emission from the hot plasma, the
electromagnetic bremsstrahlung. Even though this radiation does only play a
minor role in the energy balance, it is detectable reasonably well. That such
a diffuse x-ray radiation background exists is well known from several
measurements. But by now the proponents of the big bang cosmology interpret
it as a superposition of emission from discrete sources, which cannot be
spatially resolved with present day instruments. The reason is that in this
model there is simply no room for emission from a continuously distributed
hot plasma. But by now, in spite of large effort in this field, no discrete
sources could be detected, which can contribute considerable amounts to the
observed radiation in the spectral range beyond 10 keV.

From the equilibrium model presented here not only an accurate description of
the observed spectrum can be derived, but also the particle density coincides
well with the expected value from the calculations, applying the modified
formulas of general relativity (see appendix). The thermalised fraction
corresponds to a plasma with a mean energy of 100 keV and a density of
$4\cdot 10^{-30} \rm{g/cm^3}$. The contribution of the high energy electrons
is much lower due to the much lower collision probability, so that it is
difficult to estimate from the data. It is in the order of $1\cdot
10^{-30}\rm{g/cm^3}$ to $2\cdot 10^{-30}\rm{g/cm^3}$. This estimation agrees
well with the data derived from the equations of general relativity for the
total matter density. With a value of $70\rm{km/sec/Mpc}$ we get a value of
$6.4\cdot 10^{-30}\rm{g/cm^3}$, that means, a similar value as that, derived
from spectroscopy. This can be regarded as a strong confirmation of the model
and as an hint that really the overwhelming fraction of matter is contained
in the intergalactic plasma, as was expected from our considerations with
respect to energy loss by gravitational radiation.

\section{The radiation field}
In the last section we have discussed production and development of material
particles of the intergalactic medium. But the mechanism of the origin is
quite different from all ideas which are discussed in big bang theory. For
the radiation the situation is somewhat different. We know a large number of
processes, which all lead to the production of photons in various energy
ranges, beginning from gamma- and x-ray emission, as it is observed in
stellar explosions, the UV, visible, and IR radiation from stars, and finally
the radio emission in the long wavelength range of the spectrum.

All these photons expand nearly unhindered into space. If the universe has no
beginning, the essential question which must be answered is: Why does the
number of photons not grow to infinity? Why is the night sky dark? Of course,
one part of the answer is already contained in the general principle of
momentum loss by gravitational radiation. By this process all photons, which
are emitted, lose energy, they become 'red shifted' more and more. From the
viewpoint of energy balance this leads to a limiting range of the total
energy of the radiation field. The equilibrium level is fixed by the balance
between the loss by gravitational radiation and the emission processes.

But this balance, which appears evident from the viewpoint of energy, is not
trivial in the world of photons. Red shift changes the energy of the photons,
but not their number. Simply looking this would lead to an 'infrared
catastrophe': With decreasing energy the number of photons would exceed every
limit. But it is just this 'infrared catastrophe', which already led Max
Planck to his famous law of radiation and to the beginning of quantum
mechanics.

Basically it is the simple reasoning that photons are wave packets. They can
be regarded only as individual entities as long as the wave packets do not
overlap. Otherwise interference or possible extinction has to be considered.
The photon gas can no longer be regarded as a system of independent
particles, but must be considered as a quantum mechanical ensemble, in which
the filling of the possible energy levels is governed by their quantum
mechanical probability. This considerations are valid, of course, also for
the photons of the cosmic radiation field, when the wave packets of the
individual photons begin to interfere.

Thus we expect that similar as for the intergalactic matter there exists a
distribution of high energy photons and in addition a low energy background,
to which the emitted photons deliver their energy, but with an energy
distribution, which corresponds to a quantum mechanical equilibrium
distribution, that means, a spectrum of black body radiation. That this
background radiation exist, is known for more than 40 years, only the
explanation within the big bang model is completely different from the
interpretation given here.

Also in big bang theory people assume that in the beginning, prior to the
formation of material particles, a radiation field existed with a Planckian
energy distribution. Still after matter began to form in the expanding and
cooling space the radiation field was the dominating constituent and the
material particles remained in equilibrium with this field. Only with further
cooling and recombination of electrons and ions into neutral atoms
interactions were no longer sufficient to keep matter and radiation in
equilibrium.

It remains vague, however, why the mechanism, which was responsible for
keeping the photon spectrum in equilibrium prior to formation of matter,
should stop to work thereafter. Why should it stop just at the recombination
era, so that the inhomogeneities from this time are frozen out and can be
observed still today. In our understanding of black body radiation there is
no hint to an effect, which restricts the formation of equilibrium to some
special region of the spectrum or energy range. If there is a mechanism
which, independent from the existence and the special properties of material
walls, leads to an equilibrium distribution of photons in every closed space,
this mechanism should work, no matter if the cavity is the interior of the
sun, an oven or a freezer, or eventually the whole universe. The only
essential is the available energy density, which defines the temperature of
the radiation field.

It appears scarcely reasonable that, corresponding to the big bang model,
after formation of neutral matter the processes, which by then had guaranteed
the radiation equilibrium, suddenly should work no longer. Especially strange
is the argument that the fluctuations, which had been present at that time,
should be present still today, though the universe has expanded to one
thousand times its size and various structures have been formed during this
time. An equilibrium without interactions must be unstable. Every
perturbation will lead to irrecoverable deviations from equilibrium. Thus it
is scarcely intelligible, how one should be able to draw conclusions from the
deviations from equilibrium, observed today, to the state of the universe at
a much earlier time.

As has been discussed above, in our considerations we assume that the process
of equilibration works continuously, but also perturbations by interaction
with developing matter structures go on. Deviations from equilibrium may give
us information about these structures, on their present form or due to
running time of signals also about their past. But they represent no
indication of any development from some big bang to the present. To the
imprint, which is expected from interaction of the radiation field with the
matter distribution, we will come back with the discussion of structure
formation.

Here we have to discuss the question, if the mechanism of continuous red
shift and the thermalisation into a nearly homogeneous radiation background
of 2.73 K is reasonable. That means: Is the rate of production of photons by
the various emission processes compatible with the loss rate by quantum
interference?

The number of photons, which are lost from the background field can be
computed rather accurately. With a temperature of 2.73 K the energy density
of the radiation field is $6.3 \cdot 10^{-13}
\rm{erg/cm^3}$,which corresponds to a number of about 1700 photons per $\rm{cm^3}$.
Taking a value of 70 km/sec/Mpc for the Hubble constant, the loss rate is
$3.8
\cdot 10^{-15} /\rm{cm^{3}/sec}$.

Estimation of the production rate is much more difficult. If all matter in
the universe would radiate with the intensity of the sun, a matter density of
$3.8 \cdot 10^{-15} /\rm{cm^{3}/sec}$ would lead only to a production rate
$1.6 \cdot 10^{-17}/\rm{cm^{3}/sec}$. But we must keep in mind that there are
many sources of radiation, which are less spectacular than stars, but which
radiate much more effectively in relation to their mass. A dust grain of 1
{$\mu$}m radius in relation to its mass emits 50 times more energy than the
sun, even if its temperature is only 3 K, if both are regarded as black body
radiators. The number of emitted photons is higher even by a factor of $10^5$
due to the lower photon energy.

This demonstrates the difficulty to get a reasonable estimation of the
production rate of photons, as we have no exact knowledge of these low energy
processes like the emission of radiation from dust grains. It is just the
existence and distribution of dust in the intergalactic medium, of which we
have only a very vague knowledge. By principle we know several mechanisms,
which lead to the production of dust and discuss this topic more in detail in
the next section. But with respect to the radiation balance we must restrict
our considerations to the fact that at least from the order of magnitude it
appears reasonable that the diffuse background radiation field is determined
by the mechanism described here.

\section{Cosmic dust}
That space between the stars of galaxies is not empty, but filled with gas
and dust particles, is well known since long ago, and various mechanisms of
dust production appear well confirmed. Apparently the main source of dust are
stellar explosions, by which atoms of heavy elements, which have been formed
by nuclear fusion in the interior of stars or during explosion, are expelled
with the blast wave. During cooling on their way out they first form stable
molecules, which then coagulate into larger clusters and finally into
macroscopic dust grains.

The composition of the dust grains may be rather different depending on the
history of the explosion. It depends on the ratio of carbon and oxygen in the
expanding cloud and on the range, to which the formation of heavy elements
like silicon, magnesium or iron has proceeded by fusion reactions. Also the
size distribution of dust grains may differ dependent on the development of
density and temperature within the blast wave. Besides the dust content of
the medium, into which the cloud expands, may be different, supplying
different amounts of nucleation seed. Homogeneous nucleation requires much
higher supersaturation than attachment of molecules to an existing seed.

The expansion velocity of the clouds may be very different. The highest
speeds are observed in supernova explosions of type Ia. They are in the range
of $10^4$ km/sec, that means well beyond the local escape velocity of their
galaxy. Dust grains which are formed within a cloud at such high velocity,
would thus generally leave their parent galaxy, if they would move unhindered
into vacuum. But as the interstellar space is not empty, the dust grains will
be slowed down by collisions with gas atoms. The relative loss of momentum is
the higher, the smaller the grains. Only the largest grains can really leave
their galaxy. An estimation of the minimum size (see appendix) shows that for
a supernova, which explodes in an environment similar as that of the sun, the
limiting size of the dust grains is in the order of 1{$\mu$}m. That dust
grains of this size can be formed in the blast wave of a supernova also
results from simple estimations of growth rates under the conditions inside
the cloud (see appendix).

Supernova explosions represent a selective production mechanism, by which
large dust grains of the order 1{$\mu$}m are emitted into the intergalactic
medium, while smaller grains remain within the galaxy, serving there as
nucleation seed for later generations of explosions. The existence of these
small dust particles in galaxies can easily be proved by their selective
absorption properties. The light of stars, which transverses a layer of dust,
appears reddened compared to that of stars without absorption.

The existence of dust grains of the order 1{$\mu$}m, as we expect them in the
intergalactic medium, is more difficult to establish, as there is practically
no selective absorption of visible light. Only a general lowering of the
intensity of light from distant sources can be expected. To measure this
absorption, however, it is necessary to know the absolute intensity of the
source. Today the favoured standard candles for this purpose are supernovas
of type Ia. On one hand they have extremely high intensity, so that they can
be observed over very long distances, on the other hand observations of
supernovas in nearby galaxies have shown that their light curve follows a
universal pattern within close limits.

As has been confirmed by measurements of several research groups, indeed the
light of very distant supernova explosions appears considerably dimmer than
would be expected from their distance, determined by red shift. The adherents
of the big bang theory, however, ascribe this to an accelerated expansion of
the universe which, according to their view, has to be interpreted as the
action of some energy field with repulsive gravitation, the so called 'dark
energy', but without any physical explanation of its origin. The fraction of
this dark energy should amount to about $70\%$ of the total energy.

Interpreting the attenuation of the light from distant supernovas as
absorption or scattering by dust grains, the observed facts can be explained
without any unknown energy component. Sometimes it has been argued against
this interpretation that for very distant supernovas (red shift $z{>1}$) the
extinction is less than would be expected from a pure absorption model. But
this is true only, if space is considered as flat. In curved space the
effective angle of observation of very distant objects appears enlarged,
compared to Euclidean space. Considering both effects, absorption and
curvature of space, the observed data can be explained very accurately (see
appendix).

To estimate the amount of dust in the intergalactic medium, one had to know
the size distribution of the grains. If we assume that the dominant species
are spherical silicate grains of about 1{$\mu$}m, the observed absorption
would correspond to a mass density of about $2\cdot 10^{-32} \rm{g/cm^3}$,
that means, 0.3\% of the total mass density, as it was estimated at the
beginning of this chapter.

This value appears reasonable according to the observed number of supernova
explosions. As estimated from observations in neighbouring galaxies, in a
galaxy like ours there should be one explosion every thirty years. During the
life time of a galaxy, which we estimate in the order of one Hubble time, a
few percent of matter are processed by supernova explosions, while the main
fraction ends up in a central black hole.

If we further assume that a few percent of the stellar mass in an explosion
is transformed into large grains, which can leave their parent galaxy, a dust
fraction of 0.3\% in the intergalactic medium appears quite reasonable.

Due to the high initial velocity of the explosion dust cloud in the order of
$10^4$ km/sec, after leaving the galaxy, the dust grains will be slowed down
by collisions with the intergalactic plasma only slightly, so that they can
spread uniformly into space. Even the largest voids with extensions of 100Mpc
can be traversed within less than the Hubble time.

Thus it is reasonable to assume that the intergalactic medium, which is the
substratum for the formation of all structures, consists, apart from a
radiation field, not only of a hot plasma, but of a mixture of fully ionised
gas and of a dust component. Only by the combined action of these components
the observed structures can be understood.

\chapter{Structure formation}

In the last chapter we have discussed the composition of the intergalactic
me\-dium under the proposition that the medium is distributed uniformly in
space. Certainly this appears satisfied for the high energy plasma component,
as the interaction probability with other components is so weak that the mean
free path is several times the radius of the universe.

But for the thermalised component of the plasma this assumption is no longer
valid. The mean free path is still in the range of 100 Mpc, but it is small
compared to the radius of the universe. Thus we must expect that local
perturbations of homogeneity occur. This can lead to instabilities, by which
instead of a uniform distribution some other more stable configuration
develops.

\section{Gravitational instability}

Already more than one hundred years ago James Jeans has shown that
perturbations of the equilibrium between pressure and gravitational forces
can lead to instability, if the length scale of the perturbations is larger
than some critical value, which results from the balance equations (see
appendix). It should be mentioned here that the assumptions, which Jeans used
to derive his stability criterion, were - long before Einstein developed his
general theory of relativity - in contradiction to the accepted theory of
that time. Jeans assumed that in a homogeneous mass filled space, in
contradiction to the Poisson equation of Newtonian theory, there exists a
constant gravitational potential, and that perturbations of the homogeneous
matter distribution lead to perturbations of this potential and thus to
motion in the matter field. Perturbations of small wavelength lead to
periodical oscillations, but if the wavelength exceeds a certain limit, they
grow exponentially. The minimum wavelength which leads to instability, the so
called 'Jeans length' appears as a useful resource in the discussion of
gravitative instabilities.

The assumption of a constant potential in matter filled space, which appeared
as a deficit of the Jeans theory and was denoted as 'Jeans swindle'
frequently at his time, appears quite natural within the scope of general
relativity and as an anticipation of the correct description. According to
general relativity a homogeneous matter filled universe has to be considered
as a space of constant curvature. The amount of curvature is a measure of the
gravitational potential. Local perturbations of the matter distribution lead
to distortions of curvature, which are regarded as perturbations of the
gravitational potential. Thus the Jeans theory is just the correct
description by general relativity in the quasi-Newtonian approximation.

But in spite of that, the Jeans model is only of limited value for the
description of cosmic structures. It gives some hints to the conditions,
under which instabilities can occur. But as it is only a linear description
of small perturbations, it cannot give answers to the question, which
structures are formed, when the instabilities grow and if there exists some
new equilibrium state, into which they develop.

When we pose the question, which equilibrium distribution develops, when the
homogeneous distribution becomes unstable, it appears more reasonable to
presuppose the existence of instabilities and consider directly the
properties of the final new equilibrium state.

In the second chapter we have discussed the thought experiment of the
formation of a 'vacuum bubble', the formation of a void within a otherwise
homogeneous medium. We have stated there that the formation of such vacuum
bubbles is not observed under terrestrial conditions, as pressure forces due
to thermal motion of atoms exceed gravitational forces by many orders of
magnitude. Under cosmic conditions this is no longer true, however. The Jeans
criterion shows that the balance can be in favour of gravitational forces.

Within the linear ansatz of the Jeans model positive and negative deviations
from equilibrium and thus enhancement and reduction of matter density have
equal probability. But a more detailed consideration shows that the formation
of voids is much more probable. With every enhancement of density the
potential energy, which is set free, is applied to the interior in form of
kinetic energy leading to an increase of pressure, which counteracts any
further enhancement. Density reduction, on contrary, leads to transport of
energy to the outside, where the energy can be distributed into a much larger
volume. Thus the growth of bubbles can proceed unhindered.

Thus we must expect that gravitational instability leads to formation of
bubbles or voids. This development will stop only, when neighbouring bubbles
begin to influence each other, that means, when the pressure between the
bubbles increases considerably. In equilibrium the intergalactic plasma will
obtain a foamy structure with large voids, separated by filamentary
structures of higher density.

To estimate the size of structures, which are formed by this process, we can
again look at the balance between pressure and gravitational forces. A bubble
can be regarded as stable, if the mean kinetic energy of particles at the
boundary of the bubble is no longer sufficient to traverse the bubble against
the rising gravitational potential. Assuming as an idealised model a
spherical vacuum bubble within a homogeneous environment with properties, as
we have discussed for the intergalactic plasma in the previous chapter, that
means,a density of $4\cdot 10^{-30} \rm {g/cm^3}$ and a mean particle energy
of 100 keV, the size of the bubbles is of the order 150 Mpc.

To obtain a more accurate estimation one has to regard the fact, of course,
that the transition between voids and matter filled space is continuous and
not stepwise and that the density in the matter filled filaments is higher
than we have estimated from the diffuse x-ray intensity under the assumption
of a homogeneous density distribution. A detailed description would require
extensive numerical modeling. Here we will confine ourself to this crude
estimation, but discuss, how this estimation is confirmed by observation.

Direct spatially resolved measurements of the x-ray background with
sufficient accuracy are not available at present, by which inhomogeneities of
the intergalactic plasma could be made visible. But we can expect that with
the further development of cosmic x-ray diagnostics this may become possible
in the near future.

The distribution of visible matter, especially the distribution of galaxies
can be regarded, of course, also as an indication of the inhomogeneous
distribution of matter. As we will discuss more in detail in the next
section, galaxies will form only in the regions of the highest matter
density. Thus the distribution of galaxies can be regarded as kind of an
image of the total matter distribution. Indeed this image of the sky shows a
cellular structure with large voids. Most galaxies concentrate on a
filamentary structure separated by large voids in the order of 100 Mpc, just
in the range predicted by our model.

Also the microwave background can give us some hints to the structure of the
intergalactic plasma. Though the radiation field due to quantum interference
always tends to the equilibrium distribution of a black body spectrum, by
interaction with the electrically charged particles of the intergalactic
plasma it is continuously disturbed. By inelastic collisions (Compton
scattering) kinetic energy is transferred to the photon field. The effective
temperature of the microwave radiation thus appears a little bit enhanced
after passing through a matter concentration. According to the big bang
theory this mechanism, well known as Sunyaev-Zel'dovich effect, is of minor
importance. Only the high plasma concentrations in galaxy clusters should
exhibit an observable effect. The existence of an intergalactic plasma, which
contains the overwhelming fraction of the total mass of the universe, changes
this situation, of course.

The probability that a photon is scattered by an electron is not very high.
Under the proposition that the size and density of the universe correspond to
the values discussed here the probability that a photon is scattered during
surrounding the total universe one time is about 20\%. But by now the
microwave background has been measured with such high accuracy that even
inhomogeneities in the order of $10^{-5}$ can be detected. It it customary to
express these inhomogeneities by evaluation of the correlation function with
respect to spherical harmonics. The strongest deviation from an homogeneous
distribution are found at a angular correlation of $2\pi /220$. Applying a
radius of the universe of 4200 Mpc, as it results from the model presented
here, this corresponds to a correlation length of 100 Mpc. This is just the
same order as the diameter of the vacuum bubbles, as we have estimated from
our simple stability model.

Thus it appears reasonable to assume that the observed intensity fluctuations
of the microwave background are an indication of the inhomogeneous
distribution of the intergalactic plasma. The explanation, given by the
proponents of the big bang model, appear somewhat questionable in comparison.
According to this model the fluctuations should be formed at a very early
time, when the radiation field was still coupled to the matter field. But the
source of the fluctuations remains unclear.

\section{Thermal instability}
From the process of structure formation by gravitational instability,
discussed in the last section, we can understand the large scale distribution
of matter, but the balance of loss by gravitational radiation and the
introduction of energy remains unchanged by these structures, as this balance
is nearly independent from the distribution of matter.

But with increasing density an additional loss mechanism occurs:
electromagnetic bremsstrahlung. For a plasma with density and temperature
corresponding to the mean values of the intergalactic gas this loss amounts
only to a small fraction of the loss by gravitational radiation (see
appendix), but is has the effect that regions, in which the density was
somewhat higher than the mean, cool off faster compared to those of lower
density.

By this mechanism the pressure in the denser regions is reduced, so that more
gas can flow into these regions of overdensity. The increase of density
further enhances the radiation loss, so that the pressure and temperature
fluctuations are increased. As in this case the consequence of increasing
density is decreasing pressure, contrary to the gravitational instability,
now local concentration is preferred compared to density reduction. Thus
inside the denser filaments of the gravitationally formed foamy
over-structure a clumpy structure is generated by thermal instability.

The size of the overdense regions may be very different in this case. It
depends on the fact, if they develop in the narrow wall region between two
vacuum bubbles or in the vertex region with more that two adjacent bubbles.
The temporal development of the clumps is determined by the fact that not
only the thermal energy of the inflowing gas has to be radiated away, but
also the temperature enhancement must be compensated, which is generated by
transformation of potential to kinetic energy.

Due to the opposite effects of pressure enhancement by concentration and
pressure decrease by radiation the formation of this structure is a slow
process. It is not an instability in the actual sense. The gas is always
close to thermodynamic equilibrium, but tries to adapt to the slowly changing
conditions. As long as the density fluctuations are on a length scale, which
is small compared to that of gravitational instability, the influence of
gravitational forces is weak compared to gas kinetics. Changes are nearly
isobaric. That means, the density increases in the same manner as the
pressure drops by radiation loss.

To maintain constant pressure there must be a constant inflow of matter,
however. This mechanism can work, of course, only if sufficient matter is
available and if the inflow does not require a higher velocity than that of
free fall.

At the beginning of contraction these conditions are always satisfied. But
with increasing density the energy loss is accelerated, so that the time
constant of free fall may exceed that of radiative cooling. When the inflow
of matter is no longer sufficient to sustain pressure equilibrium, the system
becomes unstable. The plasma clouds are disrupted into smaller pieces. This
process of fragmentation is again determined by the balance of pressure and
gravitational force. But contrary to the large scale formation of voids in
this case the formation of matter concentrations is preferred, as also here
the pressure decreases with rising density.

The individual matter clumps will further shrink inside the intergalactic
plasma, while the cooling goes on. As the individual parts are in motion
during fragmentation, they will remain in motion against each other
afterwards, while their interior development proceeds autonomously. The
different velocities and the mutual gravitational attraction may lead,
however, to collisions of individual clouds. They may melt together again, or
may gain angular momentum by the action of tidal forces.

\section{Formation of galaxies}

At the end of the last section we have discussed the mechanism, which leads
to fragmentation of cooling plasma clouds. As long as the dominant energy
loss rate is electromagnetic bremsstrahlung, this process proceeds rather
slowly. But as soon as the plasma has cooled so far that recombination into
neutral atoms begins, that means, at mean particle energies in the order of
10 eV, the cooling process by recombination radiation and line radiation runs
so fast that a massive fragmentation starts.

To get an idea, how large the individual fragments will be, we can again take
advantage of the Jeans' stability criterion. Also this fragmentation is
controlled by the balance of pressure and gravitational forces. But now
enhancement of density does not produce a pressure increase, as this is
compensated by increasing radiation loss. Again overdense regions are
preferred to the formation of bubbles. Under the assumption that
fragmentation starts at about 10 eV mean energy and that up to this point the
development has been isobaric, that means, the product of temperature and
density has not changed much, the stability limit of fragmentation can be
estimated. Based on the density and temperature values, which have been
derived from the diffuse x-ray background (energy 100 keV, density $4\cdot
10^{-30} \rm{g/cm^3}$, see appendix) the length scale resulting from the
Jeans criterion is 1.7 Mpc, which corresponds to the typical distances
between galaxies. The total mass contained inside a volume of this size with
$2.5 \cdot 10^{43} \rm{g}$ is about that of a small galaxy.

This is only a very crude estimation, of course, as it is based on the
assumption of constant density and temperature inside the cloud. Besides an
undisturbed evaluation of the medium has been assumed from the formation of
the primary structure up to the beginning of recombination. but the
estimation shows that gravitational instability together with radiative
cooling leads to the formation of plasma clouds, which can be regarded as the
progenitors of galaxies.

From the existence of a stability limit it can be understood, why galaxies
cannot have arbitrary size. 'Thermal instability' is independent of the size
of the underlying region and there can be matter concentrations of very
different size. But the instability, which occurs with the beginning
recombination, leads to fragmentation into individual clouds with a rather
narrow size distribution.

By this mechanism we can understand that galaxies can develop as individuals,
in groups, or even in clusters with thousands of members, but that the size
of the galaxies is similar in all cases.

What we have denoted as galaxies are only the progenitors of the strongly
radiating objects which we observe in the sky, of course. Even recombination
and line radiation is to weak to be observed from very distant objects.
Besides the time span of recombination is only a short episode compared to
the total development of a galaxy, so that the chance is low to catch one
just in this stage. While the time span from injection of high energy
particles into the intergalactic medium to the formation of a thermal plasma
is a multiple of the Hubble time and the cooling phase, until recombination
starts, is in the order of one Hubble time, recombination is terminated
within a small fraction of this time.

With the end of recombination also the emission of electromagnetic radiation
comes to an end nearly completely. In the beginning there is a rest from
vibrational transitions of hydrogen molecules, but also this mechanism can
only slowly reduce the temperature. The existence of clouds, formed from
neutral hydrogen, is well known from the observation of the absorption of
radiation of background sources (mainly quasars) in the spectral range of the
hydrogen resonant lines (the Lyman-alpha forest). Proponents of the big bang
theory are still searching for an explanation, how such a cloud of neutral
hydrogen and helium can develop into a galaxy.

The model described here delivers a plausible explanation, if we take into
account that the intergalactic plasma besides of ions and electrons contains
considerable amounts of dust grains. During the concentration phase of the
hot plasma the motion of these grains follows the contraction of the plasma
clouds. Collisions transfer energy to the dust, which is then emitted as
infrared radiation from the surface. The temperature of the grains remains so
low, however, that even in a very hot environment evaporation does scarcely
occur (see appendix).

In the hot plasma radiation from dust grains is negligible in the energy
balance. But this changes at the end of the recombination phase, when
gravitational radiation is the only competing process. Under these conditions
radiation from dust grains becomes the dominant loss mechanism, which is
effective even at very low temperatures, as long as this is higher than that
of the background radiation field.

This cooling is accompanied by a second effect, for which the dust grains are
responsible, too, and which is essential for the further development of
galxies.In a hot plasma the collision rate between gas atoms and dust grains
is so high that the dust grains, apart from a slow Brownian motion, follow
the general motion of the plasma. With decreasing temperature collisions with
gas atoms become less frequent, so that dust grains can practically move like
free particles. They feel their mutual gravitational attraction, forming a
nearly pressureless gas, which again obeys the stability criteria of
gravitation. Dust grains coagulate into larger clusters and then into clouds
of increasing size, a process, as it is assumed in a similar way in big bang
theories for the dark matter ('hierarchical clustering'). Only when the dust
clouds have reached a sufficient size, they can attract also more and more of
the surrounding gas and thus initiate the formation of stars.

Here we will not go into the details of the development of stars, but only
emphasise the differences, compared to big bang theory. The primary
difference is related to the question, how the first stars in a galaxy can be
formed. According to big bang theory the medium, from which the galaxies
formed, consisted only of hydrogen and helium. But there is no accepted
theory, which can describe, how only from these gases condensations can be
formed, which finally develop into stars. Also all observations show that
regions, which exhibit strong formation of stars, contain much dust. For
stars, which form in later development stages of galaxies, it is not a
problem to explain the presence of dust by emission from explosions of
earlier generation stars. But for the first generation of stars this
explanation cannot be used.

According to the ideas of our model, which are confirmed by the absorption of
light from distant supernova explosions, the gas, from which galaxies are
made, already contains dust with a mass fraction of about 0.3\%, which, as
described above, coagulates and is available as condensation seed for the
first generation of stars. An essential difference between the formation of
the first and later generation of stars is, that in the beginning dust is
present in form of larger grains. Later also the gas contains the elements,
which have been formed in stellar explosions (the 'metals') as well in form
of individual atoms and very small grains.

While in the formation of the first stars first the dust grains coagulate and
the actual star of hydrogen and helium forms around this kernel, the gas of
later generation stars is enriched with metals, so that the characteristic
radiation of these elements is visible in the spectrum.

The observed poor metallicity of the oldest stars in the galaxy is not an
indication of little metal content and thus for a formation from a pure
hydrogen and helium atmosphere. The metals are only hidden in the inner
kernel, which has been surrounded by a metal poor envelop.

There is no reason to assume that the observed low metal content of old stars
is an indication of some general development of the universe with time, as it
is claimed by big bang theory. On contrary the steadily improving
observations of the last years point to the fact that even very far galaxies
posses a metal content similar to the nearby ones. This shows that also these
galaxies apparently contain young and old stars as well.

\section{The life of galaxies}
While the formation of galaxies appears as a rather straight-line process,
their further development is determined by a large number of different
processes running on different time scales. In addition the relativ motion of
galaxies with respect to each other may lead to collisions, by which they may
merge or at least exchange part of their matter or angular momentum. The
initial conditions for the further development may vary considerably. Varying
angular momentum of the protogalaxy apparently is one of the most significant
properties, which on a longer time scale determines the different modes of
development. On the angular momentum it depends, if the matter distribution
remains more or less elliptical, in which the gravitational equilibrium is
sustained mainly by the proper motion of stars, or if the contraction
perpendicular to the axis of rotation is slowed down by the centrifugal
force, so that a flat disk of a spiral galaxy is formed.

Most of the processes, which occur during the life of a galaxy or a cluster
of galaxies, for instance the formation of stars, their development and their
different stages of life, and finally their final fate by explosion or as a
neutron star or white dwarf, do not differ much in the model presented here
from what is generally believed. We will not go into the details of these
processes, but only highlight a few points, where the interpretation of
observed phenomena is different.

In the last chapter one essential difference has already been mentioned: the
existence of dust in the intergalactic medium, which enables the formation of
the first stars in a young galaxy.

The second important difference is the continuous new production of hot
plasma, by which the development of a galaxy or a galaxy cluster cannot be
regarded as completely isolated from the rest of the universe. Hot gas can
continuously flow in from outside or high energy particles can be captured.

The third difference is that gravitational radiation has to be considered
with the consequence that on the scale of the Hubble time there exists
practically no stationary motion. All motions relativ to the cosmic 'inertial
system' must slow down. Kinetic energy is reduced on account of the
gravitational potential.

\section{Galaxy clusters}
Let us first look at the hot plasma in clusters of galaxies. The diffuse
emission of x-rays from the space between galaxies shows that this space must
be filled with hot plasma. The emitted spectra can be explained as
bremsstrahlung from a fully ionised plasma with a kinetic temperature in the
order of 1 to 10 keV and particle densities up to $10^{-2}\rm{cm}^{-3}$.
According to big bang theory this was initially cold gas, which has gained
energy while flowing into an existing dip of the gravitational potential. But
as also metal lines are observed in the spectrum (the metallicity is of the
same order as in the solar spectrum), a large fraction of this gas should
already have passed a cycle of star formation and supernova explosion. But
the observation of very far clusters shows that this state must have been
reached very shortly after the big bang. With increasing knowledge of the
details of distant clusters this interpretation becomes more and more
improbable.

If we assume, on contrary, that the universe is all over filled with an
intergalactic plasma and that this plasma is regenerated continuously, it can
easily be understood that during and after the formation of a galaxy hot
matter streams in from outside. By the increasing concentration the energy
loss by radiation increases, so that not only the energy gain from the
gravitational potential is compensated, but that there is an additional
cooling compared to the intergalactic plasma.

Also the observed high metallicity of the cluster plasma can easily be
understood. As has been discussed in the last chapter, the metallicity of the
intergalactic plasma is high, even if the metals cannot be observed, as they
are bound in larger dust grains. But in the cluster gas, where the density is
higher by a factor of 1000, the erosion of the grains is strong (see
appendix), so that spectral lines of metals can be observed in the gas, too.

From the intensity of the bremsstrahlung we can try to estimate the matter
density of the cluster gas. But if we assume that the plasma is an ideal gas
in thermodynamic equilibrium, the estimated matter density is far too low to
keep the system of galaxies and plasma in balance. The observed velocities of
proper motion of the galaxies are far too high to constitute a
gravitationally bound system. A five to ten times higher total mass would be
required to avoid exceeding the escape velocity.

According to the big bang theory the problem is solved by the assumption,
that the overwhelming part of the cluster mass is present in form of so
called 'cold dark matter (CDM)', some stuff which interacts with other matter
only by gravitation, but otherwise remains completely undetectable. But apart
from the fact that up to now there exists no plausible idea, what this 'dark
matter' could be, also the overall picture is not consistent in itself.

From model calculations within the CDM model one finds that the density of a
matter cloud, which is subject only to gravitational forces, in a spherical
cluster should decrease radially as $1/r^3$. But the observed intensity
profiles of x-ray emission follow, apart from a narrow central region, a
rather exact power law all over the observed range. But for a gas in
thermodynamic equilibrium such a profile is compatible only with a matter
distribution, which decreases as $1/r^2$, unless the temperature of the gas
changes extremely (see appendix).

Also our model does not immediately give an explanation for the 'missing
mass', but we can find another interpretation of the observed facts, which
makes the search for CDM unnecessary.

Let us start from the assumption that a galaxy cluster consists only of the
galaxies and the intracluster gas, which has been concentrated and cooled
from the intergalactic plasma. We find immediately that the radial profile of
the cluster gas, which results from a thermodynamic equilibrium model,
exactly agrees with the profile derived from x-ray emission, if we assume
that the plasma temperature shows a slight increase from the centre to the
outer region (see appendix). This increase appears plausible, considering the
fact that the cluster is embedded into the much hotter intergalactic medium.
An increase of the temperature is observationally found for practically all
clusters at least for the inner region.

In the outer regions the various measurements differ considerably. Some
authors find constant or decreasing temperatures, others find increasing
temperatures. But the error margin of these measurements is very large, as
most of them are carried out in the spectral range $<10\rm{keV}$, where the
dependence of the intensity distribution on the temperature distribution is
weak. In addition the sensitivity of the detectors decreases strongly with
increasing energy. For this reason it is scarcely possible with a signal,
which consists only of a few photons, to subtract exactly the background
resulting from the intergalactic plasma, when this background is higher than
the cluster radiation itself. Already an error of 10\% of the background
correction may change an increasing temperature profile into a decreasing
one.

But what is the reason for the discrepancy between the densities determined
from the x-ray emission and those, which result from gravitational
equilibrium relations? The main reason is the assumption that the cluster
plasma can be described using the thermodynamic equilibrium relations of an
ideal gas. Reality is quite different from that. Observations with high
spatial resolution of clusters in our 'neighbourhood' (e.g. the Coma cluster)
show strong density and temperature fluctuation on a length scale of 1 to 10
kpc. This is an indication that the intracluster plasma is not homogeneous,
but exhibits a cloudy structure. On the other hand the gas is permeated by
turbulent magnetic fields, which change the transport mechanisms.

That the cluster plasma cannot be distributed homogeneously is immediately
intelligible, considering the fact that galaxies move through the gas with
high velocities and, according to their gravitational attraction, drag with
them part of the gas, until by interaction with the next mass concentration,
parts of the cloud are separated again. With time this mechanism of
continuous drag and separation generates a turbulent system of gas clouds,
which move against each other with similar velocities as that of the
galaxies. That means that a large fraction of the kinetic energy of the gas
is not present as thermal energy of individual particles, but as directed
motion of larger clouds.

Directed motion within a cloud does not contribute to collisions and thus to
emission of bremsstrahlung. What we observe, is the radiation, which is
emitted, when plasma clouds traverse each other. The energy spectrum of the
observed radiation in this case is determined by the relative speed of the
clouds. If this motion is directly coupled to the motion of galaxies, it
becomes clear, why we observe a rather exact proportionality between the
velocity dispersion of galaxies and the temperature of the x-ray emission of
the cluster gas. But as bremsstrahlung is emitted only, if gas clouds
traverse each other, we can understand, why the intensity of x-ray emission
is lower than in a homogeneous plasma. At present our knowledge of the
details of this mechanism is too poor to make exact statements on the
reduction factor.

The second point, as can be supposed, is immediately connected to the first
one. Detailed measurement on radio emission of clusters indicate that they
are permeated by magnetic fields. The strength of these fields is in the
range of 1 to 10 $\mu G$, and they change direction and strength on different
length scales from 1kpc to 100 kpc. Only part of these fields is correlated
to the motion of galaxies. For the small scale changes it is supposed that
they are generated directly inside the plasma. Regarding the fact that the
plasma is not at rest but in turbulent motion of individual clouds, the
generation of magnetic fields appears as a natural consequence.

Even if the strength of the magnetic fields in only in the range of $\mu G $,
it has a strong influence on the transport properties in the plasma. The fact
that charged particles can move freely only in the direction of the field,
while in perpendicular direction they gyrate around the field lines,
transport in perpendicular direction is strongly reduced, compared to an
ideal gas. Not only thermal conduction, the energy transport by electrons, is
reduced, but also momentum transport by ions takes place practically only in
direction of the field lines. Pressure gets a tensorial character with
different components parallel and perpendicular to the field lines. In a
turbulent plasma with statistical distribution of the field direction the
mean of pressure gradients is reduced by a factor of three, compared to a
plasma without strong fields.

That the effect of magnetic fields is so strong, we can easily get into
consciousness, comparing the cluster plasma with the magnetised plasmas,
which are used on earth in nuclear fusion experiments. The temperatures of
fusion plasmas are similar as those of the cluster gas. The magnetic fields
used for confinement are higher by about ten orders of magnitude, but also
the particle densities are higher by about twelve orders. Deviation from
isotropic behaviour are thus expected to be much higher in the cluster plasma
than in fusion plasmas. The radius of gyration is smaller than the mean free
path by many orders of magnitude for electrons and ions as well.

For the thermodynamic equilibrium in galaxy clusters that means that the
effective pressure gradients are lower by a factor of three, compared to the
values one would expect in an ideal gas. Thus the equilibrium between
pressure gradients and gravitational forces requires only one third of the
mass.

Together with the reduction of x-ray emission due to the partly ordered
macroscopic motion of plasma clouds the reduction of the required mass
density, compared to an ideal gas in thermodynamic equilibrium, allows an
explanation of the observations without the necessity of a dark matter
component.

We should emphasise here that the reduced density in the central part of
clusters does not imply that the total mass is smaller, too. The definition
of a 'total mass' is somewhat questionable, anyway, as clusters are not
isolated system in a vacuum environment, but are embedded into the
intergalactic plasma. Compared to the model calculations, by which the big
bang theory tries to ascribe the matter distribution to some 'dark matter',
the density profiles according to our model are much shallower ($1/r^2$
instead of $1/r^3$ for large $r$). Thus the aberration effects of light from
distant background sources (gravitational lensing), which are based on the
integrated mass distribution along the line of sight, may give similar
results for both theories, even if the density in the centre of clusters is
strongly different.

\section{Spiral galaxies}
Dependent on the size of the initial matter concentrations, which are caused
by gravitational instability of the intergalactic plasma, galaxies may be
formed as individuals, in groups or in large clusters. But to enable the
transformation of a matter concentration into a galaxy with luminous stars,
according to our knowledge first individual parts must cool down well below
the ionisation limit, so that again on a smaller scale gravitational forces
can exceed the pressure forces and the cold matter can contract into
individual stars. During this process individual protostars or groups of
stars may develop proper motions, which finally lead to a stable state of the
galaxy and prevent the matter from collapsing immediately into the centre of
the concentration. The stars form kind of a gas, in which the proper motion
can be regarded as a pressure, which counteracts gravitational attraction.

But this simple picture is changed, if the protogalaxy possesses a reasonable
amount of angular momentum. By the centrifugal force the contraction
perpendicular to the axis of rotation is hindered and the matter cloud
develops into a more or less flat disc, in which equilibrium is maintained
more by the balance of gravitation and centrifugal force than by the proper
motion, the 'pressure force' of stars. Only near the centre the pressure
force dominates. Thus a spherical kernel, the 'bulge', is formed, surrounded
by a flat disc, in which again regions of higher and lower star density
develop due to gravitational attraction. The matter cloud develops into the
typical image of a spiral galaxy.

But why it is a spiral? If we look at other systems, which are stabilised by
the balance of centrifugal force and gravitation, like the motion of planets
around the sun or the rotation of moons or dust belts around the planets, we
always see that the orbits are approximately circular. This is easy to
understand, as circular orbits are the most stable ones against perturbations
by neighbouring masses. Why is the picture different with rotating galaxies?

The answer is in the nature of gravity. As has been explained in the first
chapters, every motion in the universe feels a retardation by emission of
gravitational radiation due to the finite velocity of gravitational
interaction, the back reaction of the motion on the metric of space. On the
time scale of the Hubble time there exists no true stationary motion. Also
with respect to angular momentum there is no 'conservative system' except of
the total universe. Only as the orbital time of the planets or moons is very
short compared to the Hubble time, we do not recognizee the change of angular
momentum. The effect would by principle be measurable within the lunar
ranging program, which allows to measure the orbit of the moon with
centimetre accuracy. But there are other perturbations of the orbit, which
are not known with sufficient accuracy to separate them from possible
gravitational retardation.

The orbiting time of galaxies is not so short, however, that the loss of
angular momentum during on revolution could be neglected. For the sun the
loss of angular momentum in one revolution about the centre of the Milky Way
is about 2\%. As a consequence circular motion is replaced by spiral motion.

How strong the spiral structure develops, depends on the initial amount of
angular momentum, which is transferred to the system at the time of
fragmentation. Of course angular momentum can be changed also later by tidal
interaction or collision with other galaxies. Thus the range of different
galaxy types is very broad, ranging from ellipticals with very low angular
momentum to spirals with very extended arms. Formation of new stars takes
place mainly in the arms, while old stars gather in the central bulge. But on
the time scale of the Hubble time all stars move in the direction of the
centre on spiral orbits.

Also within the disc the formation of stars is not homogeneous. Gravitational
attraction leads to concentration in individual filaments with higher density
of stars with dark spaces between them. Interaction between the stars leads
to proper motion of stars within the spiral arms, but the velocity remains
low compared to the rotational motion of the disk. Only when the spiral
motion approaches the central bulge, the density of stars becomes so high
that the mutual interaction changes the rotational motion more into an
isotropic distribution, as mentioned at the beginning of this section. The
total angular momentum can change, however, only to the extent of loss by
gravitational radiation.

\section{Active galaxies and quasars}
What happens, when more and more matter concentrates in the centre of a
galaxy, we have already shortly discussed in the second chapter. As the
repulsive force of the potential energy with its negative pressure term
becomes important, contrary to the ideas of the big bang model, matter does
not collapse into a 'black hole' in the sense of a singularity. Under ideal
conditions instead an equilibrium configuration would occur, in which the
gravitational attraction of matter is in balance with the repulsive action of
its own potential energy.

But in reality we cannot expect that the transition into this equilibrium
state is completely symmetrical. Normally the matter flowing into the core
has a certain amount of angular momentum, so that we must expect that the
'black hole', as we will call it here, too, will rotate very rapidly. With
increasing concentration the rotation slows down due to increasing
gravitational radiation, as in addition to gravitational bremsstrahlung also
quadrupole radiation becomes important. But we must expect that the increase
of angular momentum from the inflow of matter exceeds this loss by far, so
that the result is a rapidly spinning object.

Also the magnetic fields, which are coupled to the infalling stars and gas
clouds, will compensate only by part, so that, if magnetic flux is conserved,
the high concentration within a very small volume will lead to extremely high
field strength.

With respect to the accretion of matter from the surrounding the scenery in
our view does not differ much from conventional theory, only that it contains
an additional simple explanation for the loss of angular momentum, which is
necessary to understand, why matter really falls into the centre of mass. The
conventional theory assumes that there exists some energy dissipation by
internal friction, that means, that there is a kind of viscosity of galactic
matter, which slows down the rotation. Of course, this possibility exists,
but the loss of angular momentum by gravitational radiation is sufficient to
explain the accretion.

An essential difference to conventional models occurs only in those regions,
where the gravitational action of potential energy is comparable to that of
the matter itself. In the idealised case of a spherical matter distribution
this state would be reached at the Schwarzschild radius. The real collapse in
the centre of a galaxy is more complex, however. Due to rotation and radial
motion of matter towards the centre and in addition also by the energy of the
magnetic field the balance between attractive and repulsive forces is far
from spherical symmetry.

While in the plane of rotation due to the kinetic pressure of inflowing
matter may preponderate, at the same time in the direction of the rotation
axis the repulsive action of the negative pressure from potential energy may
already dominate. The dynamic equilibrium in this case is characterised by a
continuous inflow near the plane of rotation and a simultaneous outflow along
the rotational axis. But all the matter goes through an intermediate state,
in which the potential energy is of the same order of magnitude as the
equivalent energy of its rest mass. During the process of accretion and
ejection the energy loss by radiation is negligible. The process can be
regarded as adiabatic. Thus the amount of kinetic energy will attain the same
order as potential energy.

The intermediate state within the 'black hole' is characterised by a mean
energy of the heavy particles or a 'temperature' corresponding to their rest
mass energy. This energy is far beyond the binding energy in the atoms, so
that the identity of individual atomic species is completely lost. The state
of matter is comparable to that, which is called a quark-gluon-plasma by high
energy physicists, or, if you want, to the state, which the proponents of the
big bang theory assume as the state of matter at the time of elementary
synthesis. During the ejection from the 'black hole' all unstable particles
will decay. Protons and neutrons will combine into stable light nuclei by
part in a similar way as described by conventional models of elementary
synthesis.

But dependent on the history of a black hole the conditions may be somewhat
different, so that from the ratio of the elements we cannot draw conclusions
on the global density of the universe or other parameters like that. Contrary
to the picture of big bang theory elementary synthesis takes place still
today, so that eventually higher concentrations of some unstable nuclei are
expected, for which the time since the big bang was too long to survive.

Characteristic of the collapse in the core of an 'active galaxy' is on one
hand the intensive radiation from the compact accretion disk and on the other
hand the outflow of matter in high energy plasma beams, the 'jets' in the
direction of the rotation axis. The extremely narrow collimation of the jets
can be explained by the fact that the emission originates in a region of
extremely high magnetic field strength. The motion of the ejected matter,
which consists of electrically charged particles, is thus coupled very
strongly to the direction of the field lines. Thus most of the particles are
guided back into the accretion disk along the closed field lines.

Only in the direction of the magnetic poles matter can leave the region of
very high density and field strength. So it formes those strongly collimated
jets, which consist of fully ionised 'primordial matter', that means, of
protons, helium nuclei, and electrons, which leave the core with high
velocity. Due to the conditions at the origin the kinetic energy of the jet
should be of the same magnitude as the energy equivalent of the heavy
particles, the velocities should be relativistic.

Direct observation of the jets appears difficult, as emission of radiation
from fully ionised matter is weak. The overwhelming fraction of kinetic
energy is contained in directional motion of the outward flow, internal
interaction inside the jet is comparatively poor. Only if the jet collides
with matter at rest in the vicinity, intensive interactions take place.
Direct collisions lead to radiation in the gamma and x-ray spectral range,
secondary effects produce optical and radio emission. As the matter fields,
which are traversed by the jets, normally are not homogeneous, but consist of
individual clouds or in some cases of parts of neighbouring galaxies, there
occur luminous knots, which are dragged along with the jet by momentum
transfer. Far in the outer region, where the energy transferred to the
environment is distributed over a larger area, large clouds are observed,
from which intense radio emission is observed.

From the linear structure of the jets out to very large distances we must
conclude that only a small fraction of the jet energy is lost by interaction
with surrounding matter. The main fraction will be distributed into the
universe without further interaction. So it is available as the seed of hot
matter of primordial composition, from which new structures can be built.
Thus the cycle is closed from the source of matter in the black holes and
their jets along with the formation of more and more complex structures up to
the collapse of galaxies and the generation of new sources.

\chapter{The universe - a zero-sum game?}
In the last chapters we have demonstrated that it is possible to develop a
model of the universe, which is based only on known and experimentally
testable physical principles. With this model we can understand the
astrophysical observations without applying any unknown or unprovable new
effects or constituents.

It is true that also this model brings us no step closer to the answer of the
question, why the basic laws of nature are as they are. By principle they
might be arbitrarily complex and the universe might contain any number of
components, which can by no means experienced by us. There could be, as some
physicists believe, many different universes with different laws and
different history. But if these universes do not influence our universe at
all, the question of their existence is completely irrelevant for a
physicist.

The advantage of the model presented here is, compared to all these
considerations, that to understand the cosmic phenomena we rely only on
mechanisms, which are, at least by principle, experimentally testable. In
addition this model contains no singularities, in contrast to conventional
big bang theory. The well proved principles of general relativity are not
questioned, but only supplemented. The questions of causality, homogeneity
and structure formation are solved quasi automatically without invoking any
new 'godhead' like 'dark energy' or 'inflation'.

The singularities of the big bang and of black hole formation, which are
mathematically describable, but physically scarcely acceptable, do not occur
in the model presented here. Also the attempt to construct a quantum theory
of gravitation, becomes superfluous. Matter states, in which gravitational
interaction would become so strong that they require a quantum mechanical
description, that means, interactions on the scale of the Planck length, do
not occur. The repulsive action of potential energy prevents every quantum
from penetrating this range.

In addition the attractiveness of the model is enhanced by the fact that, as
already Einstein has emphasised, the size of the universe is finite, if, in
contrast to the big bang model, the curvature of the universe is positive.
The conception of an infinite space, which continuously expands, and for some
unexplained reason even with accelerating velocity, is much more difficult to
imagine than a closed, but unlimited continuum, as in this case we know the
two-dimensional analog in form of the surface of a sphere.

Also the total mass and the energy content of the universe are finite
quantities. Introduction of potential energy, which gives a negative
contribution to the total energy content, if we interpret the invariant trace
of the energy tensor as the 'total energy', lead to the very simple result
that the total energy of the universe is zero. Also quantum fluctuations do
not contribute to the energy balance, as the virtual states add the same
amount to the positive energy as they add to the negative potential energy.
This fact immediately solves the riddle, why the curvature of space is so
many orders of magnitude less than quantum mechanics predicts.

But in contrast to Einstein's static universe, in which he tried to establish
a balance between the attractive and repulsive forces by introduction of the
'cosmological constant', the equilibrium described here is stable, as the
repulsive action increases with increasing curvature and density.

Of course we can put the question: Why is the quantity of mass in the
universe just so that there exists equilibrium between potential energy and
rest energy of matter? Couldn't we think of a universe, in which there exists
only the energy of the quantum vacuum? If we trust the Einstein equations,
the answer is: no. These equations lead to contradiction, unless one allows
the generation of matter or radiation fields. If on the other side one
assumes that matter can be produced, the solution of the time dependent
equations always runs into an equilibrium state, in which the gravitational
action of quantum fluctuations vanishes.

Considerations, how such a process of matter generation could look like, are
highly speculative, of course. According to our present view of high energy
and quantum physics no process is known, which leads to the generation of
matter. Also if the energy remains in balance due to the simultaneous
generation of mass and potential energy, there remains the problem that
generation of real particles from vacuum fluctuations always leads to
creation of pairs of matter and antimatter particles. But the world, which we
know, practically contains only matter, that means leptons and hadrons, but
only negligible amounts of the corresponding antiparticles.

But this fact can scarcely attributed to our understanding of gravitation.
From the insight that the total energy of the universe balances to zero it
appears more reasonable to suspect that our view of the world of elementary
particles is not quite realistic. Maybe that the sum of all elementary
particles and their antiparticles is zero indeed, but we have overlooked this
symmetry by now.

The proposition that hadrons are composed of three quarks and not of
antiquarks and the assumption that quarks and leptons are genuinely different
particles makes a symmetrical theory practically impossible. Only if the
elementary building stones of electrons and protons are equal, a symmetrical
concept can be realised. Below we will delineate the speculative attempt, to
develop such a symmetrical model of elementary particles, which at least
describes the basic observed phenomena correctly, but without the claim of
representing a confirmed theory. But it may contain a new way of thinking,
which could lead to a revision of our the actual standard model.

One basic idea of this speculative model is the assumption that the
elementary particles, from which matter is built, are neither bosons nor
fermions, but that the spin of particles results from different coupling
between these elementary fields. Also in the current theory electrons are
described by a system of coupled field functions, which are coupled into a
spinor field. The properties related to spin result from this special kind of
coupling, which takes place in some abstract 'spin-space', however.

The idea of the model presented here is that the elementary fields - let us
call them quarks again for simplicity - are spatial fields, which can
interact in different ways: On on hand by superposition in form of a spinor,
forming a pointlike system like an electron, on the other hand as spatially
extended systems, comparable to the system of proton and electron in a
hydrogen atom or the binding between quarks and antiquarks in mesons.
Contrary to the 'fermionic' states of spinors such 'bosonic' bound states are
possible with different levels of energy or angular momentum.

Four propositions are sufficient to describe the spectrum of known elementary
particles in such a symmetrical model:

1. Every particle contains the equal numbers of quarks and antiquarks.

2. Quarks and antiquarks can be bound together 'fermionic' (as a spinor,
pointlike, spin quantum number 1/2) or 'bosonic' (spatially extended, integer
spin)

3. Quarks have an electrical charge (the elementary charge) or hypercharge
(strangeness) or both, antiquarks the same, but with opposite sign.

4. No two quarks can be in the same quantum state.

Equal to the basic triplet in conventional theory there are three different
quarks and the corresponding antiquarks. Thus the spectrum of mesons, the
bosonic bond between one quark and an antiquark is identical. The spectrum of
leptons results from the corresponding fermionic binding. But in this case
there is only one quantum state for every combination of quarks. No excited
states or states with different angular momentum are possible (for details
see appendix).

With the families of mesons and leptons the possibilities of two-quark states
are exhausted. But there is a number of different realisations of states
consisting of two quarks and two antiquarks. Hadrons can be interpreted as a
system of two fermionically bound quarks, surrounded by two bosonically bound
antiquarks. also in this case we find the same spectrum as in conventional
theory. As the 'classical' quarks form a complete group, the same quantum
states would be obtained, if we replace one quark by the sum of the two other
antiquarks. Only the spin properties would be different. This is compensated
in our model by the fact that there is one fermionic and two bosonic bonds.

Systems consisting of two fermionically bound quarks, bosonically bound to a
pair of fermionically bound antiquarks, form an additional system of very
heavy mesons (J/$\psi$ or Y), W- and Z-particles can be interpreted as
fermionically bound states of two quarks and two antiquarks. Details of the
spectrum, which result under consideration of the condition of
distinguishability (point 4) are discussed in the appendix.

As in conventional theory interactions between elementary particles can be
described as exchange reactions. But the processes of strong and weak
interaction must now not be regarded as completely different mechanisms. Both
have the effect that one quark is replaced by another one. The only
difference is that strong interaction relates to a bosonic bond, while in
weak interaction a fermionic bond is changed.

It is interesting to remark that in the scheme presented here electron and
proton together contain one quark and one antiquark of all three species. The
same holds for neutron and electron neutrino. Thus it appears plausible that
the universe contains equal amounts of quarks and antiquarks of every type.
Thus there exists the possibility that the universe could be created from
nothing.

Of course it remains to explain, why the separation into protons and
electrons is stable, though the sum of their quarks is equal to the sum of
the antiquarks, that means, in total equivalent to zero. That they cannot
annihilate each other becomes clear regarding the fact that interactions of
electrons and protons take place either with the bosonically bound antiquarks
or with the fermionic core built from two quarks. In both cases the total
number of the involved quarks is different from the number of antiquarks, so
that annihilation is impossible. Only when the density of matter is so high
that the bosonic bonds become unstable, that means, e.g. in the interior of
black holes, creation or annihilation can be expected.

The concept, which we have outlined here, requires a radical change of
thinking, of course. The world is no longer distributed into fermions and
bosons and the components of wave functions no longer describe superpositions
of possible states, but the superposition of real fields. But maybe that just
here we need a radically new ansatz. With quantum theory we have the
mathematical tools to describe the phenomena of the microcosmos in an
excellent way. But the price is that the underlying physical reality is
scarcely intelligible for our imagination.

Maybe that the world on the smallest and largest scales becomes easier to
understand, if we are willing to leave the accustomed tracks in spite of the
impressing success in the last century. Already for many times the ideas of
mankind about the structure of the world, though appearing well confirmed,
have been radically changed by new thoughts and new directions of view. This
paper should be regarded as an impulse to extend our angle of view.

\begin{appendix}

\chapter{The definition of density and  potential energy}
Basis of the theory of general relativity are the Einstein equations
\footnote{In the mathematical formulations we follow the notation of the
standard textbook of R. M. Wald 'General Relativity', The University of
Chicago Press, Chicago (1984)} , which relate the metric of space-time,
expressed by the curvature tensor (or its contracted form, the Ricci tensor
$R_{ij}$) and the energy tensor $T_{ij}$.

\begin{equation}
\label{einst}
R_{ij}\,-\frac 1 2 R \times g_{ij}\,=\,8\pi T_{ij}
\end{equation}
(in a unit system, where the speed of light and the gravitational constant
have been set to 1). $g_{ij}$ is the fundamental tensor. The first and second
derivatives of this tensor determine the curvature tensor. The condition of
general covariance, that means, the requirement that the content of the field
equations should be independent of the choice of the reference system, leads
to certain restrictions for the possible forms of the energy tensor. It must
be determined uniquely by scalar or tensorial quantities, which depend only
on local invariant properties of matter itself, on the fundamental tensor or
on quantities which can be derived from this tensor without differentiation.

In the description of cosmological problems usually the energy tensor is used
in the form borrowed from special relativity. It is described as an ideal
fluid medium with the total energy expressed by a continuous matter density,
with its 4-velocity, and by the pressure tensor.

\begin{equation}
T_{ij}\,=\,\rho \frac{dx_i}{ds}\frac{dx_j}{ds}\,+\,p \,\left(g_{ij} +
\frac{dx_i}{ds}\frac{dx_j}{ds}\right)
\end{equation}

But in general relativity the definition of density is no longer unique. In
Euclidean space the density is defined as the differential of the mass $\rho
= dm / dV$. If space is curved, however, the size of the spatial volume
element depends on the metric. According to Einstein therefore the energy
tensor used in his equations should contain the invariant density, that
means, the density, which one would measure in an Euclidean space which is
locally tangential to the actual space. The quantity $\rho _0 = dm / dV_0$
($dV_0$ is the volume element in tangent space) is a local property of the
matter field and thus can be used in a covariant equation.

But the use of the invariant density leads to the problem that the total mass
is no longer the result of integration of this density over the actual
volume.
\begin{equation}
\int{dm} \,\ne\,\int{\rho_0 dV}
\end{equation}

By principle it is impossible to define mass density and total mass as
invariant properties, if the volume element is not invariant. To obtain
invariance of the total mass, we have to introduce a corrective term in the
energy tensor. The density in curved space can be expressed by the invariant
density as
\begin{equation}
\label{rho0}
\rho\, = \, \rho_0 \, + \,\rho_0 \left ( \frac{dV_0}{dV} -1\right ).
\end{equation}
Only the second term in this equation depends on the metric. It is a scalar
property, which can be used in a covariant energy tensor, but not together
with some vectorial quantity like the 4-velocity. The only reasonable method
to change the energy tensor in such a way that the matter term, the quantity
$T_{00}$, is equal to $\rho$, and that in spite of that the property of
covariance is preserved, is to follow the way proposed by Einstein, but to
introduce an additional term, which allows for the influence of curvature:
\begin{equation}
\label{Tij}
T_{ij}\,=\,\rho _0 \frac{dx_i}{ds}\frac{dx_j}{ds}\, + \,\rho_0 \left (
\frac{dV_0}{dV} -1\right )\times g_{ij}\,+\,p \,\left(g_{ij} +
\frac{dx_i}{ds}\frac{dx_j}{ds}\right)
\end{equation}
The meaning of the additional term can easily be recognised, looking at the
Schwarzschild metric, which describes the space-time outside a radially
symmetric matter distribution with the total mass $M$. This metric is
characterised by the path element
\begin{equation}
ds^2 \,=\,-f(r)dt^2 \,+\,h(r)dr^2\,+\, r^2 d\Omega ^2
\end{equation}
with $h(r)=1/(1-2M/r)$ ($d\Omega$ is the element of solid angle). From this
we get the volume element
\begin{equation}
\label{dV}
 dV \,=\,\frac{r^2drd\Omega}{\sqrt{1-2M/r}},
\end{equation}
from which in the limiting case of weak curvature we find the density
\begin{equation}
\rho \,=\,\rho_0\,-\,\frac {\rho _0 M}{r}.
\end{equation}
The difference compared to the density in tangent space is just the density
of potential energy in the sense of Newtonian mechanics. To maintain
conservation of matter or total energy not only locally but also integrally
in curved space, it is necessary to include potential energy in the energy
tensor.

The path used here to calculate the potential energy is not applicable
generally, of course. Due to the special choice of coordinates the volume
element (\ref{dV}) is singular at $r=2M$, so that the expression in
(\ref{rho0}) is no longer meaningful.

Equation (\ref{Tij}) is valid also inside a matter distribution. But as the
size of the volume element depends implicitly on the metric, in the
determination of the metric from the Einstein equations (\ref{einst}) we have
to consider the reactive coupling of the metric to the energy tensor. In the
discussion by now we have implicitly assumed that only the mass density
determines the metric of space. If there are additional kinetic contributions
like pressure in a gas, instead of matter density we must use the density of
total energy, expressed by the invariant trace of the energy tensor.

The exact determination of the potential energy may thus become rather
intricate. But we can make some general statement, how the solution should
look like:

1. Potential energy is proportional to the local density $\rho _ 0$ and thus
vanishes in empty space.

2. As a positively curved volume element is always larger than the
corresponding element in the flat tangent space, the quantity $dV_0/dV -1$ is
always negative.

3. By principle potential energy can attain arbitrarily high negative values,
if the size of the curved volume element (like in the limit of the
Schwarzschild metric) becomes singular.

4. The local energy density (trace of the energy tensor) can be positive or
negative. By the pressure term in the energy tensor potential energy acts
like a negativ pressure or repulsive force.

Based on these properties of potential energy, formation of singularities in
the density distribution, the formation of 'black holes' is impossible. In
the conventional view of general relativity kinetic pressure is regarded as
an additional positiv contribution to the total energy and thus enhances the
attractive action of gravitation. Thus stabilisation of a system by pressure
gradients becomes impossible beyond a certain mass limit. On contrary
potential energy generates a negative pressure and thus a repulsive action,
increasing with density, and thus brings every contraction to an halt.

\newpage
\chapter{Basic equations for a homogeneous universe}

Based on the assumption that the universe does not contain any regions with
distinct properties but is homogeneous on large scale, it can be described by
a global metric of the form
\begin{equation}
ds^2 \, = \,-dt^2 +a(t)^2\cdot[d\psi ^2\,+\,f(\psi)\cdot(d\theta^2+\sin
^2\theta \,d\phi^2)].
\end{equation}
Here one assumes that the metric can vary in time, but is spatially
homogeneous. The function $f(\psi)$ takes the values $\sin^2\psi$, $\psi^2$,
or $\sinh^2\psi$ for positive, zero ore negative curvature respectively. Here
we restrict our discussion to the case of positive curvature.

In the case of a homogeneous matter distribution with density $\rho$ and
pressure $p$ the solution of the Einstein field equations leads to the well
known Friedmann equations, to determine the temporal evolution of the scale
parameter $a(t))$:
\begin{equation}
\label{fried1}
\frac{3\dot{a}^2}{a^2} +\frac{3}{a^2}= 8\pi\rho
\end{equation}
\begin{equation}
\label{fried2}
\frac{3\ddot{a}}{a} = -4\pi(\rho + 3p)
\end{equation}
(in a unit system with the speed of light and the constant of gravity set to
1). If there are various fields with different equations of state. that
means, different relations between $\rho_i$ and $p_i$, the right hand sides
of (\ref{fried1}) and (\ref{fried2}) have to be replaced by the corresponding
sum. The equations (\ref{fried1}) and (\ref{fried2}) can be satisfied
simultaneously, however, only if density and pressure vary in a suitable
manner with change of the scale parameter $a(t)$. A pressureless matter field
requires $\rho_m\propto 1/a^3$, a radiation field with $p=1/3\,\rho_r$ has to
change as $\rho_r\propto 1/a^4$. Global conservation thus exists only for
matter. A radiation field changes its total energy on account of a change of
the scale variable.

If we postulate global conservation for all kinds of fields, we have to take
into account potential energy. To include potential energy in the energy
tensor, any reasonable formula must contain besides of the fundamental tensor
only invariant local state variables, the local invariant density of the
energy fields and the local curvature of space. Potential energy must vanish,
if this density is zero or if the gravitational potential and with this the
curvature of space is zero. As the most reasonable ansatz we must consider an
expression of the form $\lambda \times g_{ij}$. But in contrast to the
'cosmological constant' introduced in Einstein's static solution, now the
quantity $\lambda$ depends on the density of matter and on the spatial
curvature. In analogy to Newtonian mechanics the best choice for $\lambda$ is
an expression of the form
\begin{equation}
\lambda \,=\,\frac{K}{a} \sum_k \left(\rho_k+3p_k\right)\,
\end{equation}
where $K$ is a universal constant, which is related to the gravitational
constant. This dependency of $\lambda$ on density and curvature has far
reaching consequences, however:

1. The system (\ref{fried1}) and (\ref{fried2}) no longer allows a time
dependent solution, but only the trivial solution $a=const.$. A homogeneous
universe necessarily has to be static. The curvature is fixed by the
requirement that the attractive force of positive energy is in balance with
the repulsive action of potential energy. In a universe, where pressureless
matter dominates, we have
\begin{equation}
\frac{3}{a^2}= 8\pi\rho-\frac K a \rho
\end{equation}
\begin{equation}
0 = -4\pi\rho - \frac K a \rho,
\end{equation}
leading to
\begin{equation}
K=-4\pi a,\quad a = 1/\sqrt{4 \pi \rho} \,\quad\rm{oder} \,\quad a/\rm{cm}=
3.28\cdot 10^{13}/\sqrt{\rho /\rm{cm}^{-3}}
\end{equation}

2. Vacuum energy from quantum fluctuations, as they have been discussed as
possible origin of a cosmological constant in conventional theory and which
are described by energy fields of the form $C\times g_{ij}$ are without any
effect in the field equations, as their contribution to potential energy is
just $-C\times g_{ij}$ and thus compensates the field energy exactly.

3. Local changes of fields with zero rest mass like kinetic energy or
radiation fields generate changes of curvature and thus changes of the scale
parameter $a$ proportional to $a^{-4}$. At the same time the potential energy
of matter is changed with the same proportionality, so that the total energy
of the system remains constant. Like in Newtonian theory there is no law of
conservation for kinetic or potential energy alone, but only for their sum.
\newpage
\chapter{Red shift - energy loss by gravitational radiation}
Apart from the geometric interpretation of gravitation the essential
difference between Newtonian mechanics and general relativity is the
statement that gravitational interaction and thus any change of space-time is
limited to the speed of light. As a consequence, similar to electromagnetic
interaction, due to the retardation of potentials there exist radiation
effects. This gravitational radiation is not restricted to the quadrupole
radiation, which reduces the angular momentum of rotating double stars, but
also in the 'force free' or geodesic motion in matter filled space there
exists a momentum loss by gravitational radiation, analogous to
bremsstrahlung in electromagnetism.

An exact derivation of the effect from the field equation is not known by
now. A simple estimation can be obtained, however, from a quasi-Newtonian
approximation, only considering the retardation of interaction and the
curvature of space. To do this we calculate the force on a moving test mass
in a homogeneous matter-filled universe under the following assumptions:
\begin{enumerate}

\item
The velocity of gravitational interaction is limited to the speed of light.

\item
Mass and energy create an intrinsic curvature of space. A spatially
homogeneous universe can thus be regarded as a three-dimensional surface of
constant curvature.

\item
Lines of force between masses follow the geodesic lines between the
corresponding points.
\end{enumerate}

To determine the motion within a three-dimensional surface it is convenient,
to describe it by means of polar coordinates of a four-dimensional sphere of
a given radius $a$.
\begin{eqnarray}
\label{coord}
x=r \cos \chi \cos \theta \cos \phi &,&y=r \cos \chi \cos \theta \sin
\phi ,\nonumber \\z=r \cos \chi \sin \theta &, &w= r \sin \chi,
\end{eqnarray}
where $x,y,z$ and $w$ is a set of Cartesian coordinates. In this system we
determine the force exerted on a mass at the point $P=(a,0,0,0)$ by the mass
in some volume element $dV$ of the surface at $r=a$.

The size of the volume element is given by
\begin{equation}
\label{vol}
dV= a^3 \cos ^2 \chi \cos \theta \:d\chi \:d\theta \:d\phi
\end{equation}
The distance between the element and $P$ , as measured along a straight line
through the four-dimensional space is given by
\begin{equation}
\label{rp}
r_P = \sqrt{(x-a)^2+y^2+z^2+w^2}=a\sqrt{2-2\cos \chi \cos \theta \cos
\phi}.
\end{equation}
The distance along the connecting geodesic line is
\begin{equation}
\label{s}
s= 2a \arcsin \frac {r_P}{2a}=2a \arcsin \sqrt{\frac{1-\cos \chi \cos
\theta \cos \phi}{2}}
\end{equation}
The component of the gravitational force at $P$ in some direction, defined by
the unit vector $\vec e$ then is
\begin{equation}
\label{dF}
dF=\frac{G\rho m dV}{s^2} (\vec{e_s} \vec e ),
\end{equation}
where $\rho$ is the mass density, $G$ the gravitational constant and $\vec
{e_s}$ the unit vector in direction $s$ at point $P$. As all directions are
equivalent, we chose $\vec e$ in the direction of the y-coordinate. The
projection onto the direction of the force component is
\begin{equation}
\label{ee}
(\vec{e_s} \vec e )= \frac {\cos \chi \cos\theta \sin \phi}{ \sqrt{1-\cos
^2\chi \cos ^2 \theta \cos ^2 \phi}}.
\end{equation}
In a universe of constant mass density we find the total force by integration
of Eq.(\ref{dF}) over all directions and distances:
\begin{equation}
\label{Fy}
F_y = G \rho m a^3 \int _{-\infty} ^{\infty}\int _{-\pi /2}^{\pi /2}
\int _{-\pi /2}^{\pi /2}
 \frac {\cos ^3\chi  \cos ^2\theta  \sin \phi}{ s^2 \sqrt{1-\cos ^2
\chi\cos ^2 \theta \cos ^2 \phi}} \,d\chi \,d\theta \,d\phi
\end{equation}

The limits of integration over $\phi$ are set to $\pm \infty$, as the lines
of force may extend beyond the reciprocal pole.

Obviously for a mass at rest the integral is zero on symmetry grounds. In
Newtonian mechanics this holds also for a moving mass, as the gravitational
force acts in every volume element simultaneously. If interaction is limited
to the speed of light, however, retarded forces have to be used. In the
determination of the force all distances $s$ at time $t$ have to be replaced
by $s'(t) = s(t-\tau)$, their value at a point of time, earlier by the
running time of the signal $\tau = s'/c$. That means that the integral is no
longer symmetrical. Distances in direction of motion appear enlarged,
opposite to motion they are reduced. This leads to a resulting force opposite
to the direction of motion.

For the numerical evaluation of the force integral it is convenient to change
the coordinate system. Instead of changing the distance $s$ and its
projection in Eq.(\ref{Fy}), it is easier to keep these quantities unchanged
and instead to change the integration variable, by introducing a comoving
reference system:
\begin{equation}
\label{phi}
\phi '= \phi (t')= \phi (t)+ \frac{d\phi}{dt} \tau =  \phi (t)+ \frac{vs'}{ac},
\end{equation}
where $v$ is the velocity of the moving mass. This is allowed, as $\phi$ is
the only coordinate, which is affected by the motion. Expressing the distance
$s$ according to Eq.(\ref{s}) the differential reads:
\begin{equation}
\label{dphi}
d\phi = d\phi '\left ( 1-\frac{v}{c}\frac {\cos \chi \cos\theta \sin
\phi '}{\sqrt{1-\cos ^2\chi \cos ^2 \theta \cos ^2 \phi '}}\right).
\end{equation}
The limits of integration are not changed by the transformation. The
resulting force then is
\begin{eqnarray}
\label{Fy2}
F_y &=& -2G \rho m a^3 \frac {v}{c}\int _{-\infty} ^{\infty}\int _{-\pi
/2}^{\pi /2}
\int _{-\pi /2}^{\pi /2} \times \nonumber\\ &\times&
 \frac {\cos ^4  \chi\cos ^3\theta \sin^2 \phi ' \:d\chi \:d\theta \:d\phi '}{ (1-\cos ^2
\chi \cos ^2\theta \cos ^2 \phi ')\left(\arcsin\sqrt{(1-\cos \chi \cos \theta
\cos \phi ')/2}\right)^2}
\end{eqnarray}
From numerical integration of the integral we find the value $Y=3.0695$. Thus
the change of velocity results as
\begin{equation}
\label{dv}
F_y = m \frac{d v}{dt}= - mv \frac{2Y \rho Ga}{c}.
\end{equation}
Correspondingly we find for the change of kinetic energy
\begin{equation}
\label{dE}
\frac{d E}{dt}= \frac{d }{dt}\frac{mv^2}{2} = -\frac{4Y \rho Ga}{c} E.
\end{equation}
Every motion in mass-filled space is subject to a loss of kinetic energy by
emission of gravitational radiation, caused by the finite velocity of
gravitational interaction. Here we have calculated the energy loss only for
the case of non-relativistic motion, but as the formula for the energy loss
does not contain the mass explicitly, it appears reasonable to assume that it
is valid also for relativistic motion, in the limiting case also for energy
quanta without rest mass. That means that photons experience the
corresponding energy loss. Of course in this case the energy loss cannot
change the velocity, but instead the wavelength will change.
\begin{equation}
\label{lambda}
\lambda =\lambda _0 \exp \left( \frac{4Y \rho Ga}{c}t \right ).
\end{equation}
As a first approximation this leads to a linear increase of wavelength with
distance, exactly the same behaviour as described by big bang theory and as
it is confirmed by observations. Differences between the two theories occur
on one side, when deviations of the exponential dependence from linear
approximation become important, and on the other side when the different
assumptions on the temporal development of expansion are taken into account.

We should emphasise that for the case of Euclidean space the formula leads to
similar difficulties as discussed in the well known Olbers paradox. As
according to Eq.(\ref{lambda}) red shift is proportional to the radius of the
universe, in this case the red shift would be infinite. This can be regarded
as an indication that the size of the universe must be finite.

The quantity $(4Y \rho Ga/c)$ in Eq.(\ref{lambda}) is identical with the
Hubble constant $H$. On the other hand from the Einstein field equations we
have the relation
\begin{equation}
\label{R}
a=\sqrt{\frac{c^2}{4 \pi G \rho}}
\end{equation}
(for better clarity, contrary to the last chapter, we have used the
traditional writing with the speed of light $c$ and the constant of gravity
$G$) This leads to the relation
\begin{eqnarray}
\label{all}
\rho=\frac{\pi H^2}{4Y^2G},&a=\frac{\displaystyle{cY}}{\displaystyle{\pi H}},
 & M=\frac{\displaystyle{Yc^3}}{\displaystyle{2GH}}.
\end{eqnarray}

$M$ is the total mass of the universe. Assuming a value of 70 km/sec/Mpc for
the Hubble constant, the result for the total density is $6.43\cdot
10^{-30}\rm{g/cm^3}$, the radius of the universe is $1.29\cdot
10^{28}\rm{cm}$, and the total mass is $2.73\cdot 10^{56}\rm{g}$.
\newpage
\chapter{X-ray emission of the intergalactic plasma}
We can prove the existence of the intergalactic plasma and determine its
properties from the diffuse x-ray background radiation, if we assume that
this plasma is responsible for the emission and not, as assumed in the
conventional big bang model, some unresolved individual sources. For the high
energy fraction ($ E > 3\rm{keV}$) this assumption appears justified, as
there are no known cosmic processes, which could contribute to the background
in quantities comparable to observations.

The emitted spectrum of the x-ray background has been measured with
sufficient accuracy to find out the properties of the intergalactic plasma.
In a plasma of electron density $n_e$, the ion density $n_i$, and kinetic
temperature $kT$ the energy emitted per time and frequency interval by
bremsstrahlung is given by
\begin{equation}
\label{epsilon}
\frac{d\varepsilon}{d\nu}=\frac{2^5 \pi e^6}{3m_ec^3}\left(\frac{2\pi}{3 m_ekT}\right)^{1/2}
n_e \sum n_i Z_i^2 g \,e^{\displaystyle{-\frac{h\nu}{kT}}}.
\end{equation}
$Z_i$ is the charge of the ion species $i$, $g$ the Gaunt factor, for which
we use the approximation $g=0.74(E/kT)^{-0.4}$ in the
following\footnote{Itoh, N. et al., Astrophys.J.Suppl. 128 (2000)}. For a
fully ionised plasma, consisting of hydrogen and 25\% helium for the energy
spectrum this leads to the formula
\begin{equation}
\label{eps}
\frac{d\varepsilon}{dE}=7.216\cdot 10^{-26}\rm{cm^{-3}s^{-1}}\sqrt{\frac{1keV}{kT}} \left(\frac{E}{kT}
\right)^{-0.4}\left(\frac{\rho}{10^{-29}\rm{gcm^{-3} }}\right)^2\,e^{\displaystyle{-\frac{E}{kT}}}.
\end{equation}
The intensity of radiation then results by integration along the optical
path. Here we must regard the fact that the energy changes by red shift along
the path. To determine the intensity at some energy $E_0$ at a distance $s$
we must use the emissivity at the energy $E=E_0\times exp(Hs/c)$. In addition
the intensity may be reduced by absorption from dust grains.
\begin{equation}
\frac{di}{dE}=\frac{1}{4\pi}\int_0^\infty\frac{d\varepsilon}{dE}\,e^{\displaystyle{-\kappa s}}ds.
\end{equation}
In this equation the absorption coefficient $\kappa$ has been assumed as
independent of frequency. Integration along the path can then be replaced by
integration over energies. From $E=E_0\times exp(Hs/c)$ we get $ds=c/(HE)dE$.
With the abbreviation x=E/(kT) we find
\begin{equation}
\frac{di}{dE}=\frac{1}{4\pi}\frac{c}{H}\int_{x_0}^\infty\frac{d\varepsilon}{dx}\,\left
(\frac{x}{x_0}\right)^{\displaystyle{\frac{\kappa c}{H}}}\frac{dx}{x},
\end{equation}
where $d\varepsilon/dx$ according to Eq.(\ref{eps}) is given as
$kT\,d\varepsilon/dE$. The numerical solution of this equation can be fitted
to the observational data \footnote{Gruber et al., Astrophys.J. 520, 124
(1999)} by suitable choice of the parameters $\rho$, $kT$ and $\kappa$ (see
figure). The accuracy of the data is not sufficient for an exact
determination of the parameters, as the shape of the spectrum varies not much
with varying absorption. The effective temperature varies between $\kappa
c/H=0$ and 0.5 from $kT=110 \rm{keV}$ to $kT=90 \rm{keV}$. The corresponding
density values are $\rho=3.6\cdot 10^{-30}\rm{g/cm^3}$ and $\rho=4.3\cdot
10^{-30}\rm{g/cm^3}$.

\epsfig{file=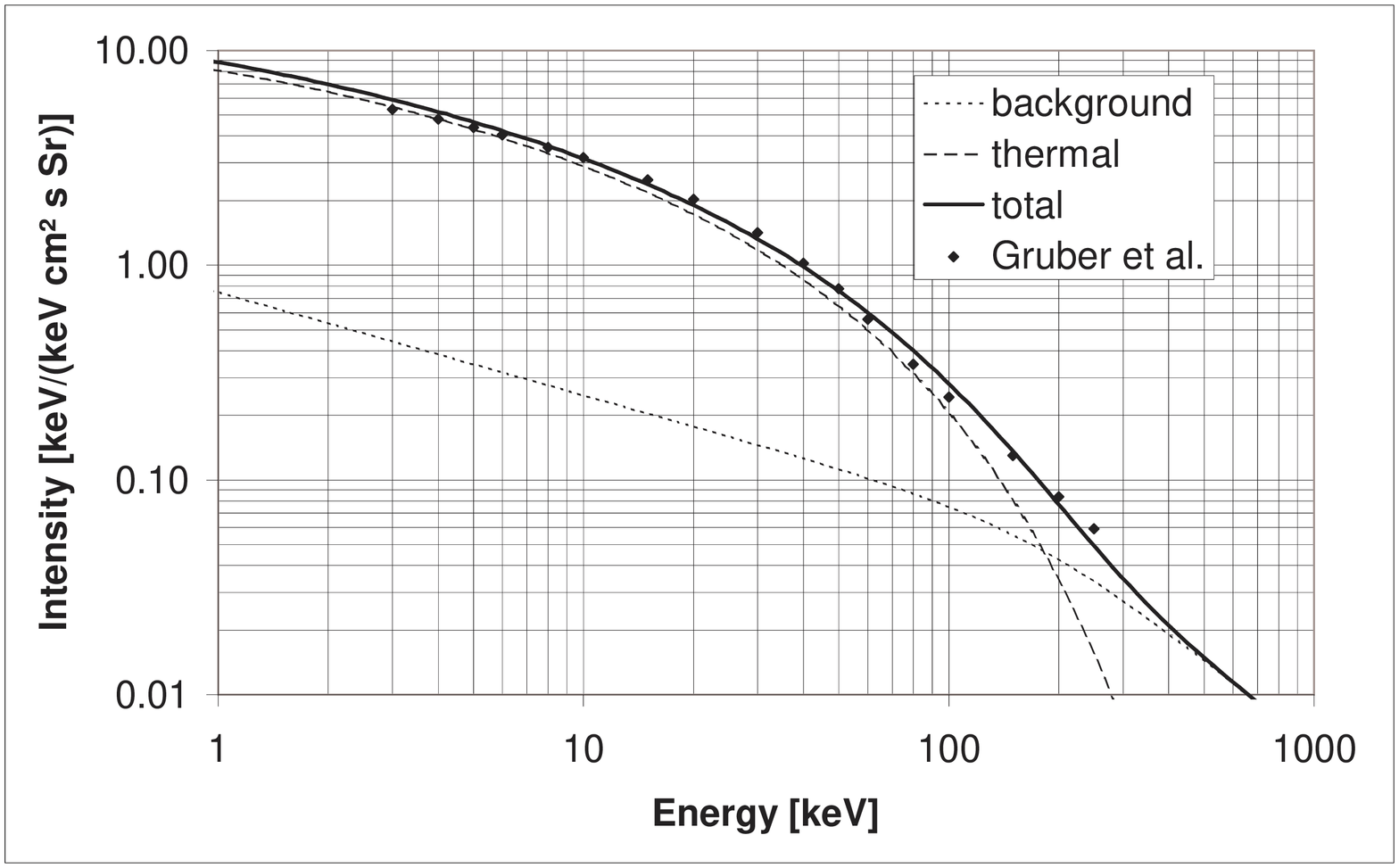,angle=-90,width=10cm}

In the range of high energies the calculated intensity distribution decreases
considerably steeper than the measured data. This is caused by the
contribution of high energy electrons. Due to the decreasing collisional
cross section the contribution of the high energy tail is considerably lower
than that of the thermal plasma.The dotted line in the figure shows the
contribution of the high energy electron under the assumption that the total
density of the high energy particles is $1\cdot10^{-30}\rm{g/cm^3}$ with an
energy distribution $dn_e/dE\sim 1/E $ between 1 GeV and 100 keV.

\newpage
\chapter{Intergalactic dust}
Formation of dust needs two conditions to be met: on one hand a relatively
high density of heavy elements and secondly a temperature range, low enough
to allow the formation of molecules, but high enough that collisions and
reactions are sufficiently frequent. To our experience these conditions are
found predominantly in the expanding clouds of stellar explosions. Preferably
supernova explosions of type Ia (SNIa) must be regarded as sources of
intergalactic dust. On one hand these explosions are characterised by very
high expansion velocity of the explosion cloud and on the other hand the
progenitor star is a 'white dwarf', a highly evolved star, thus containing a
large fraction of heavy elements.

All supernovas of type Ia show very similar properties. In the following we
consider the formation of dust in a 'standard' supernova of this
type.\footnote{Perlmutter, S. et al., ApJ 483(1997)}. SNIa occur, when a
white dwarf exceeds its stability limit (the Chandrasekar limit) by accretion
of matter from its surrounding.

\section {Formation of dust}
'White dwarfs' are stars in which the first steps of nuclear fusion, the
burning of hydrogen and helium has come to an end. Also the further fusion
reactions have practically stopped, so that the star is kept in equilibrium
with gravity only by the degeneration pressure of electrons. If the total
mass of such a star exceeds a certain limit, it becomes unstable, it
collapses. During this collapse in the centre of the star nuclear fusion is
resumed up to the final formation of the iron group atoms. This energy
production than leads to explosion of the complete star and a matter cloud,
consisting preferably of heavy elements expands radially with velocities in
the order of $10^4$ km/sec. The mass of the expanding cloud is given by the
stability limit of the star. It is about 1.4 solar masses ($M_{\rm {C}}=2.8
\cdot 10^{33} \rm g$). The 'standard' supernova reaches a peak intensity
of $M=\rm{-19.5}$, the decay time per day is $dM/dt=0.06m$. Only at the
beginning of the explosion thermal energy is produced in the centre, which is
by part emitted in form of high energy neutrinos. Later there is only
radiative energy transport in the interior and emission of radiation from the
boundary layer of the expanding cloud.

When the emission pattern and the expansion velocity are known, the
determination of the temperature in the surface layer is straightforward, if
we assume that the plasma is optically thick.
\begin{equation}
L=4 \pi R^2 \sigma T^4=10^{\frac{M_{\rm{peak}}-M}{2.5}} \times L_ {\rm{peak}}
\end{equation}
($M$ is the absolute bolometric magnitude; $ \sigma $ is the Stefan-Boltzmann
constant).

With $R=u \cdot t$ and an expansion velocity of $u=5000\,\rm{km/s}$ we find
that recombination into neutral atoms (at T=10000 K) begins about 20 days
after the intensity maximum and is completed after 50 days (4000 K). After
this time observations show deviations from the linear decrease of intensity,
as the surface layer becomes optically thin. But it appears reasonable to
assume that the bolometric magnitude continues its linear decrease and that
the observed deviations result only from the shift of emission from visible
to infrared.

With this extrapolation the equilibrium composition of the gas can be
determine under the assumption that the element abundances are similar as
those of the sun, but without hydrogen and helium and a reduced fraction of
C, N, and O, as these elements will be consumed by part by the further steps
of fusion. The numerical calculations show that the first molecules are
formed about 75 days after explosion (SiO and CO at 2500 K).

105 days after maximum intensity the temperature in the surface layer has
dropped to 1500 K. Under equilibrium conditions at this time the first
condensed phases would occur ($\rm{MgSiO}_3$ and $\rm{SiO}_2$).
($\rm{Fe}_2\rm{SiO}_4$ would solidify at 1100K).

At this time the radius of the cloud is about $4.5\cdot 10^{15}\rm{ cm}$, the
particle density is of the order $10^9\rm{cm}^{-3}$. Under these conditions
homogeneous nucleation in the gas is scarcely possible. The gas would be
strongly supersaturated. But in most cases an explosion blast wave does not
expand into vacuum, but into the interstellar medium, which already contains
small dust grains, which can serve as a seed for the further growth.

The growth of grains takes place preferably shortly after the gas reaches
supersaturation, as later the number of collisions decreases with the third
power of the cloud radius. To estimate which final size a dust grain can
achieve, we must calculate the number of collisions between a grain and the
surrounding molecules. For simplicity we assume that there are only molecules
of one species with mass $m$ and that the particle density is constant all
over the cloud $n=M_C/(4/3 \pi R^3) / m$. Then we get for the temporal change
of the size of a spherical grain of density $\rho_{\rm{D}}$:
\begin{equation}
\frac {d}{dt} \left( \frac{4}{3} \pi r^3 \rho_{\rm{D}} \right) = \pi r^2 n m v_{\rm{th}}.
\label {growth}
\end{equation}
$v_{\rm{th}}=\sqrt{3kT/m}$ is the mean thermal velocity. The time dependence
of the temperature can be approximated by a power law of the form
$T/T_{\rm{0}}=(r/r_{\rm{0}})^{\rm{-\alpha}}$, where $\rm{\alpha}$ is close to
one for the example considered here. With $R=u\cdot t$ we can integrate
Eq.(\ref{growth}) from the beginning of supersaturation to infinity with the
result
\begin{equation}
r_{\infty}=r_{\rm 0}+\frac{3}{40\rm{\pi}}\frac
{M_{\rm{C}}v_{\rm{th_0}}}{\rho_{\rm{D}}u R_{\rm0}^2}.
\end{equation}
The index 0 refers to the values at beginning of condensation. With the data
of our example we find a final radius of $2.8\rm{\mu m}$. Of course this
value can be regarded only as a very crude estimation due to the
simplifications made above. We have assumed that there is only one species of
molecules condensing at 1500 K. For another species condensing only at 800 K
there would be only half the final size.

Secondly we have supposed implicitly that the density of molecules ins not
changed by the condensation itself. If this assumption is justified, depends
on the number of seed grains, which are swept up by the expanding cloud. The
density of the interstellar medium in the vicinity of the explosion may vary
strongly. If we take the conditions in the neighbourhood of the solar system
as a guideline, the density of the interstellar medium is of the order
$10^{-23}\rm{g/cm}^3$ with a dust fraction of 1\%. Assuming a mean grain size
of $0.01\rm{\mu m}$, the density of seed grains is $10^{-7}\rm{cm}^{-3}$. The
main part of dust formation should be finished, before the explosion cloud
reaches a volume of $10^{48}\rm{cm}^3$. That means that the number of seed
grains is of the order $10^{41}$. If all matter contained in the cloud would
condense onto these grains, the radius of one grain would be $15\rm{\mu m}$.
From this estimation we can conclude that it is the collision rate and not
the availability of molecules, which determines the final size of the dust
grains. Anyhow we can state that grains, which begin growing with the onset
of supersaturation, can easily attain a size of more than $1 \rm{\mu m}$.
Seed grains, which are swept into the cloud at later times, will reach
correspondingly smaller size.

\section {Size selection}
At the time of creation the radial velocity of dust grains is equal to the
expansion velocity of the cloud. If this expansion would take place into
empty space, nearly all grains would leave the galaxy, in which they have
produced, as their velocity exceeds the local escape velocity by several
times. For instance, for a supernova, which explodes in the spiral arm of a
galaxy of say $10^{45} \rm g$ at a distance from the centre like that of our
sun (7.7 kpc), the escape velocity is about 500 km/sec, ten times less than
the radial velocity of the grains.

In real galaxies the dust grains must first traverse the interstellar medium,
before they reach free space. In every collision with a gas atom they lose
part of their initial momentum. Even if the emission is perpendicular to the
disk of a galaxy, the column density of the interstellar gas should be in the
order of $n_{\rm{c}}=10^{21}\rm{cm}^{-2}$.

In every collision with a hydrogen atom the momentum loss on average is
$-m_{\rm H}\cdot u$, as the mean radial velocity component of the gas atoms
is zero. The momentum loss of a spherical dust grain of radius $r$ and the
density $\rho_{\rm D}$ then is
\begin{equation}
\frac{4 \pi}{3} \rho_{\rm D}r^3\frac{du}{ds}=-n\cdot m_{\rm H}\cdot \pi r^2 u,
\end{equation}
or by integration along the path of the grain
\begin{equation}
u_{\rm{\infty}}=u_{\rm0}\cdot \rm{exp} [-3m_{\rm H}n_{\rm c} /(4 \rho_{\rm
D}r)].
\end{equation}
If, as has been estimated above, the initial velocity is ten times higher
than the escape velocity, the exponential factor in the equation above has to
be $> 0.1$ for a grain to leave the galaxy. For a column density of
$10^{21}\rm{cm}^{-2}$ this means $r> 2.5\rm{\mu m}$. Of course, this is only
a crude estimation of the order of magnitude. But it clearly demonstrates
that only large grains can leave the galaxy, while smaller grains are
thermalised within the galaxy and are available there as seed grains for
later generations of supernovas.

As the size limit is just in the range, where the absorption properties of
dust for optical radiation change from selective to wavelength independent
absorption, the dust, which is captured inside the galaxy has highly
selective properties and leads to reddening of spectra, while the dust, which
escapes into the intergalactic medium, exhibits 'grey' absorption.

\section {Global distribution and stability}
After leaving their parent galaxy dust grains are still exposed to the
interaction with the intergalactic plasma. This may lead to erosion or
eventually to complete destruction of the grains. To estimate the rate of
erosion, we can, similar to dust formation or momentum loss, again start from
the collision rate, but we must regard the fact that at the very high
velocity of the striking protons ($kT\approx 100 keV$) the sputtering
efficiency is low ($\eta < 10^{-2}$). Analogous to Eq.(\ref{growth}) we find
for a spherical grain with radius $r$:
\begin{equation}
\frac{d}{dt}\left( \frac{4}{3} \pi r^3 \rho_{\rm D} \right)=
-\eta m_{\rm a} n\cdot \pi r^2 v_{\rm{th}},
\label{eros}
\end{equation}
Here $m_{\rm a}$ is the mass of the eroded atom. $v_{\rm{th}}$ is the thermal
velocity of the ions in the intergalactic plasma, which is in the order of
$5\cdot 10^8\rm{cm/sec}$. By integration we get $\tau =r_0 \rho_{\rm D}/(\eta
m_{\rm a}nv_{\rm{th}})$ as the erosion time constant for grains with radius
$r_0$.

For a silicate grain of $r_0= 1 \rm{\mu m}$ with the data of the
intergalactic plasma we obtain a time constant of $>10^{18} \rm {sec}$, more
than the Hubble time. This means that intergalactic dust in the form of large
grains remains stable and is distributed homogeneously in the universe. As
the grains leave their parent galaxy with a considerable fraction of their
initial velocity in the order of some $10^3$ km/sec, they can traverse even
the largest voids of the universe during their lifetime. Thus we can expect
that the dust is distributed uniformly all over the universe.
\section{Detection of intergalactic dust}
By principle the existence of dust can be proved by absorption or emission
measurements. But due to the low intensity of the signals both measuring
methods have large error margins. This is especially true for emission
measurements. Though the dust grains are heated by collisions with the
surrounding gas particles, which lead to emission in the far infrared, their
emission is low compared to the intensity from sources in our galaxy.

An indirect hint on the ejection of dust particles from galaxies into the
intergalactic medium can be found in the properties of the gas component in
clusters of galaxies. Though also in this case a large amount of dust can
escape from the individual galaxies, the velocity is not sufficient, to
escape from the complete cluster. On the other hand the density of the
intracluster gas is higher by three orders of magnitude, compared to the
intergalactic plasma. That means that the erosion time of dust grains is only
about $10^{15}$ sec, only a small fraction of the cluster lifetime. Thus the
major part of all metals is in gaseous form, be it produced in the cluster
itself or swept in as dust from the intergalactic plasma.

Indeed the observed metallicity of the gas in most clusters is close to that
in the solar system, that means, mass fractions in the order of 1\%. This
high concentration can be understood only, if we assume that inside the
galaxies a very efficient mechanism of dust production and ejection exists.
But as the corresponding spectral lines are not observed in the intergalactic
plasma, the metals must be present there in solid form, bound into dust
grains.

The most convincing hint onto the existence of coarse grain intergalactic
dust comes from absorption measurements, though even here the interpretation
of the observations is controversial due to the lack of spectral selectivity.

Supernovas of type Ia form a very homogeneous group of phenomena and due to
their extreme luminosity are observable over large cosmic distances. But what
we observe is a stronger decrease of the radiation intensity than would be
expected from a pure distance relation ($i \sim 1/s^2$). According to big
bang theory one tries to explain this by an accelerated expansion of the
universe, caused by the action of some mysterious gravitationally repulsive
dark energy.

But absorption by intergalactic dust appears as a much more obvious
explanation. The fact that this dust in contrast to interstellar dust inside
a galaxy is much coarser, immediately explains, why this absorption does not
lead to reddening of the spectra.

In a quantitative determination of the intensity deviation from the normal
$1/s^2$ dependence of distant supernovas in addition to absorption the
influence of space curvature has to be considered. The connection between the
size of surface area and the solid angle element differs from Euclidean
geometry. In a homogeneously curved space with a radius of curvature $a$ for
a source at distance $s$ the illuminated area is smaller than in Euclidean
space by $F/F_0 = \sin^2 \varphi/\varphi^2$, with $\varphi = s/a$. The
intensity change of a point source by absorption (absorption coefficient
$\kappa$) and curvature thus results as
\begin{equation}
I=I_0\cdot e^{\displaystyle{-\kappa
s}}\left(\frac{\varphi^2}{\sin^2\varphi}\right)
\end{equation}
or, expressed by change of magnitude:
\begin{equation}
\label{dm}
\Delta m =\frac{2.5}{2.306}\left(\kappa a\varphi-2\ln\varphi+2\ln\sin\varphi\right).
\end{equation}
The relation between red shift $z$ and the angle $\varphi$ is given by
$1+z=\exp(Ha\varphi/c)$. Various observations of SNIa have been carried out
by several groups. In the figure below these measurements are plotted
together with a fit by Eq.(\ref{dm}).
\begin{figure}[hbt]
\epsfig{file=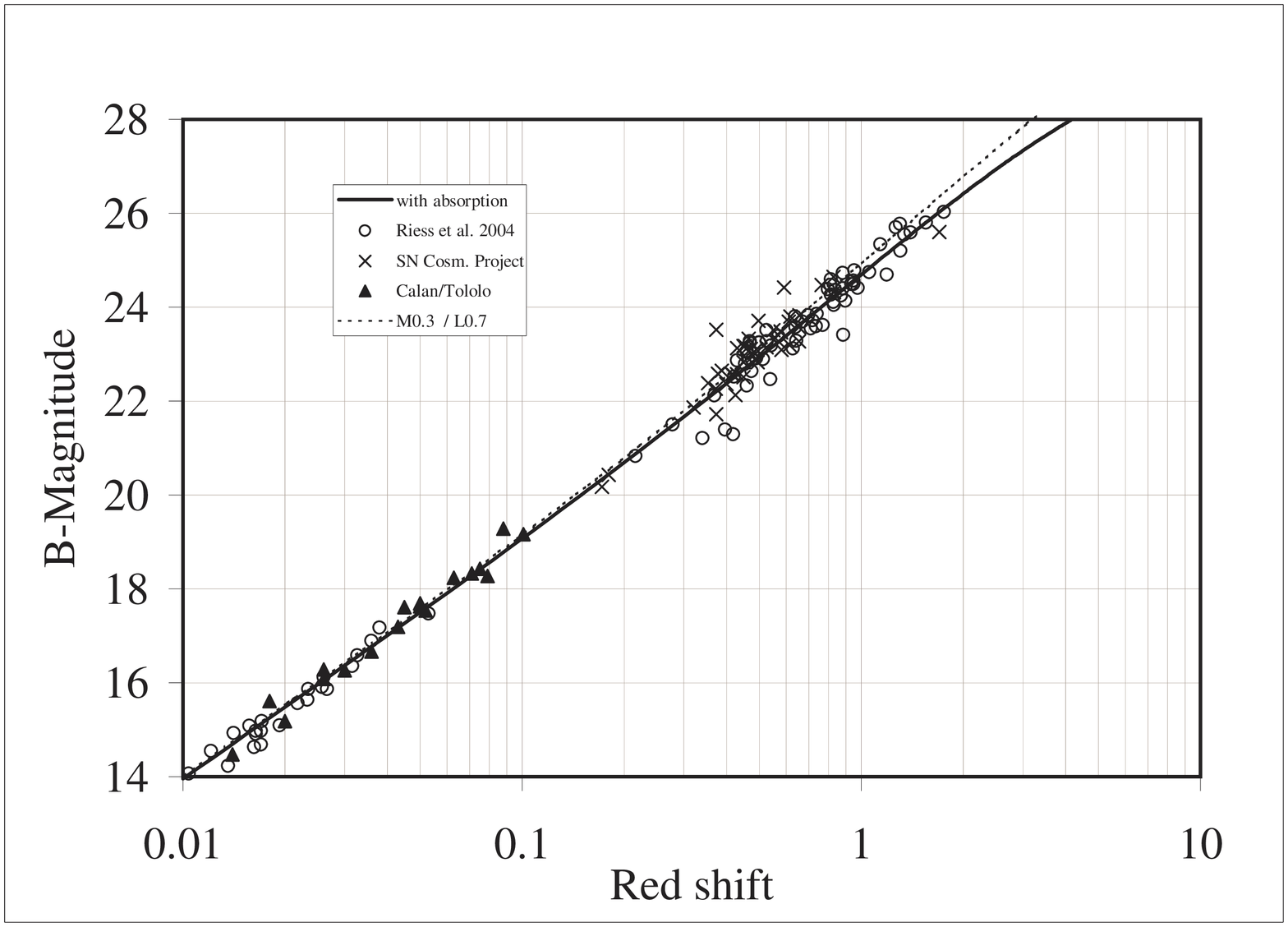,angle=-90,width=11cm}
\parbox{1cm}{$\,$  }
\parbox{11cm}{Solid line: Fit of measurements by Eq.(\ref{dm}),
dashed line: 'concordance model' $\Omega_M=0.3, \Omega_\lambda =0.7$,
measurements: crosses: Perlmutter,S. et al., ApJ 517 (1999), spheres:
Riess,A.G. et al., ApJ 607 (2004),triangles: Hamuy,M. et al., AJ 112 (1996)}
\end{figure}
In addition the result according to the 'concordance model' of big bang
theory is plotted, which ascribes the deviations from the Euclidean distance
law to an accelerated expansion of space. From the fit we find a value of
0.98 for $\kappa a$. With the value $a=1.29\cdot 10^{28} \rm{cm}$ and
$H=70\,\rm{km/sec/Mpc}$, as derived from red shift data, assuming a mean
grain radius of $1\mu m$ and a density of $2\rm{g/cm^3}$for the dust grains,
their total mass fraction in the universe is about 0.3\%, just the same order
as in the solar system and as it is confirmed by measurements of metallicity
of the gas in galaxy clusters.
\newpage
\chapter{Jeans-instability}
In the year 1902 Jeans has shown that perturbations of the equilibrium
between pressure force and gravitational attraction can lead to
instabilities. Jeans' considerations were based on the assumption that in
homogeneous mass filled space there exists a constant pressure and a constant
gravitational potential, a proposition, which at that time, more than a
decade before Einstein's theory of general relativity, was in clear
contradiction to the Poisson equation of Newtonian theory of gravitation
\begin{equation}
\Delta\Phi=4\pi G \rho_0,
\end{equation}
which was considered as generally valid and which for $\Phi=\rm{const.}$ only
allows the trivial solution $\rho_0=0$.

The reason is that the Poisson equation does not contain an energy term with
repulsive action. All the new theories contain such terms, be it Einstein's
cosmological constant, be it the kinetic term represented by the expansion of
the universe, or be it, as considered here, the negative pressure of
potential energy.

If like Jeans we start from an undisturbed matter distribution of density
$\rho_0$ and discuss the development of a perturbation of the form
$\rho=\rho_0+\rho_1({\bf{r}})$, The perturbation of the potential
$\Phi_1({\bf{r}})$ is given by
\begin{equation}
\label{phi1}
\Delta\Phi_1=4\pi G \rho_1.
\end{equation}
Together with the equation of continuity
\begin{equation}
\frac{\partial\rho_1}{\partial t}+\rho_0\nabla {\bf{u_1}} =0
\end{equation}
and the equation of motion
\begin{equation}
\label{beweg}
\rho_0 \frac{\partial {\bf{u_1}}}{\partial t} =-c_s^2\nabla\rho_1-\rho_0\nabla\Phi_1
\end{equation}
the potential equation forms a complete system to determine the quantities
$\rho_1$, $\Phi_1$ and the local flow velocity $\bf{u_1}$. $c_s$ is the
velocity of sound. For an isothermal perturbation in an ideal gas we have
$c_s^2=k_BT/m$. Considering in this system a perturbation with the Fourier
component $\exp(i\textbf{kx}-i\omega t)$, from Eqs.(\ref{phi1}-\ref{beweg})
we get
\begin{equation}
 -\omega \rho_1 + \rho_0 {\bf{ku_1}} =0,
\end{equation}
\begin{equation}
-\rho_0\omega{\bf{u_1}}=-c_s^2{\bf{k}} \rho_1-\rho_0{\bf{k}}\Phi_1,
\end{equation}
\begin{equation}
- {\bf{k}}^2\Phi_1=4\pi G\rho_1.
\end{equation}
and from this the dispersion relation
\begin{equation}
\omega^2=c_s^2(k^2-k_J^2) \quad \rm{with}\quad k_J^2=\frac{4\pi G\rho_0}{c_s^2}.
\end{equation}
For $k<k_J$ $\omega$ is imaginary. This means that the perturbation grows
exponentially. For perturbations on a characteristic length scale of more
than $l_J=2\pi/k_J$ pressure forces cannot damp out gravitational action and
the system becomes unstable. For a radially symmetric perturbation of the
matter distribution with radius $R_J=l_J/4$ and a mean density $\rho_0$ the
minimum involved mass, the so called 'Jeans mass' is
\begin{equation}
M_J=\frac{2}{3}\pi^{5/2}\left(\frac{k_B T}{Gm}\right)^{3/2}\rho_0^{-1/2}.
\end{equation}

From these formulas we find for the stability limit of the intergalactic
plasma ($k_BT=100\,\rm{keV}$, $\rho_0=4\cdot 10^{-30}\rm{g/cm^3}$)
$R_J=170\,\rm{Mpc}$, which corresponds to the largest observed netlike
structures in the universe.

With the assumption that the individual regions cool down isobaric, they
reach the recombination limit of about $k_BT=10\rm{eV}$, at which a further
fragmentation can be expected, at a density of $4\cdot 10^{-26}\rm{g/cm^3}$.
This corresponds to a stability limit of $R_J=1.7\,\rm{Mpc}$ and a mass of
$2.5\cdot 10^{43}\rm g$, the typical mass of a galaxy.

If the gas cools isobarically down to about 10 K, the density would be of the
order $10^{-22}\rm{g/cm^3}$. The critical mass in this case is $5\cdot
10^{35}\rm g$, corresponding to the mass of the largest stars, The distance
between these stars would be about $2R_J=6 \rm{pc}$, corresponding to the
observed regions of star formation. Of course this extrapolation into the
range of star formation should be looked at with caution, as here the
dynamics of the dust component is of considerable influence.
\newpage
\chapter{The mass balance of galaxy clusters}
Clusters of galaxies are the largest observable structures in the universe.
Besides of galaxies they contain large amounts of hot gas. But if we trust
the concept of the 'concordance model' the overwhelming fraction is 'dark
matter'. The necessity of a large amount of this dark matter follows from the
discrepancy between estimations of the total mass, which results from the sum
of luminous matter in galaxies and of the gas mass, radiating in x-rays, and
the mass, which appears necessary to explain, why the galaxies form a bound
system in spite of their high peculiar velocities.

Basis of most mass estimations are the properties of the cluster gas. The
density of the cluster gas can be estimated from the intensity of
bremsstrahlung (see Eq.(\ref{eps})). Under the assumption that the matter
inside the cluster is in stationary equilibrium from the balance of pressure
forces and gravitational attraction, the total mass can be estimated. For a
spherically symmetric matter distribution we have
\begin{equation}
\label{thermo}
\frac{dp}{dr}=-G\frac{\rho_{gas}}{r^2}M(r).
\end{equation}
$G$ is the gravitational constant, $M(r)$ the mass enclosed in a sphere of
radius $r$. The relation between pressure and density for an ideal gas is
given by $p=\rho_{gas}k_B T/m$. $m$ is the mean mass per gas particle, for a
fully ionised hydrogen gas with 25\% mass fraction of helium about 0.6 $m_p$
(proton mass).

Already with the application of this equation our view is different from
conventional concepts. Especially the experimental determination of the
temperature profile is influenced in many cases by the expectations of the
model upon which they are based. For the outer region the plasma temperature
can scarcely be determined spectroscopically, as the measured signals are
small compared to the background level. On the other side measurements always
give only a signal, projected through layers with varying properties.
Transforming the signals to radial profiles induces additional errors.
Besides that, most available measurements are restricted to the energy range
$<10\rm{keV}$, in which the dependence of the bremsstrahlung spectrum on
temperature is very weak, especially at high temperatures.

Conventional theories assume that the cluster gas has been introduced from
outside as cold gas, has gained its thermal energy from the changed
gravitational potential and is cooled by radiation only in the dense centre.
Thus most derivations assume a decreasing or in some cases constant
temperature in the outskirts. If we start with the assumption of a hot
intergalactic plasma, we expect a continuously increasing temperature from
centre to the outside. Measurements in the central region of most clusters
show an increasing temperature, corresponding to $T\propto\rho^{-\gamma}$
with $\gamma=0.2$ and changing to $\gamma=0.1$ at larger radii. We suppose
that this increase continues also in the region, where accurate measurements
are not possible. In contrast, conventional theory assumes values from
$\gamma=0$ (isothermal) to $\gamma=-2/3$ (adiabatic heating) for the
outskirts.

Still more important is the assumption that the plasma can be treated as an
ideal gas. Measurements of clusters in our 'vicinity' (Coma) show, however,
that the plasma is permeated by magnetic fields of the order $B=1\mu G$. That
means that the motion of charged particles is strongly coupled to the
direction of the magnetic field lines. The gyration radius $r_B=mvc/eB$ is of
the order $10^7\rm{cm}$ for electrons, but also for ions with
$10^{10}\rm{cm}$ many orders of magnitude less than the mean free path, which
in a plasma of the density $10^{-3}\rm{cm^{-3}}$ and a temperature of $10^8$K
is more than $10^{23}\rm{cm}$. For all transport phenomena only the component
of gradients in direction of the magnetic field contributes to the transport.
In a plasma, in which the coherence length of the fields is small compared to
the dimensions of the plasma, on average only in one third of the plasma the
direction of gradients is identical with the direction of the fields.
Transport rates are thus reduced to 1/3, compared to those in an ideal gas.
For energy transport this effect has been discussed in many papers. But it is
valid also for mass transport by pressure gradients, as the ions are firmly
bound to the field lines, too. Thus, compared to an ideal gas, in a
magnetised plasma we have to introduce an additional factor $\xi=1/3$ in the
balance equation (\ref{thermo}).

To determine the matter distribution of the gas and of the total mass, we
start from the measured profiles of x-ray emission. Disregarding the very
central part, where the distribution may be strongly disturbed by
concentration of gas around a large central galaxy (cD galaxy), the projected
emission profiles are well described by a so called '$\beta$ profile':
\begin{equation}
I(x)= \frac{I_0}{[1+(x/r_c)^2]^{\beta_x}}
\end{equation}
The values of $\beta_x$ normally are between 1.0 and 1.3. In this case
inversion of the projected to the radial profile gives such a '$\beta$
profile', too, but now with the exponent $\beta_r=\beta_x+1/2$. From
Eq.(\ref{eps}) thus, regarding the dependency $T\propto\rho^{-\gamma}$, we
find the expression for the density distribution of the gas:
\begin{equation}
\label{gasmass}
\rho (r)= \frac{\rho_0}{[1+(r/r_c)^2]^\alpha} \quad \rm{with} \quad  \alpha =
\frac{\beta_x+1/2}{2-\gamma /2}
\end{equation}
For some historical reason most astrophysicists use the form
\begin{equation}
\rho (r)= \frac{\rho_0}{[1+(r/r_c)^2]^\frac{3\beta}{2}}.
\end{equation}
The values of $\beta$ for $\gamma=0.2$ for this definition are between 0.53
and 0.64. For an isothermal gas the values are higher by 5\%, in the limiting
case of adiabatic heating higher by 40\%.

The density $\rho_0$ follows from integration of Eq.(\ref{eps}) over the
spectral range of the measurement and resolution of the equation with respect
to the density.
\begin{equation}
\left(\frac{\rho_0}{\rm{gcm^{-3}}} \right)^2= 7.3\cdot10^{-25}\frac{\varepsilon}
{\rm{erg/cm^3s}}/\sqrt{\frac{kT_0}{keV}}
\end{equation}
The resulting density data are higher by a factor $\sqrt{T/T_0}$ for a
temperature profile, increasing to the outside, than for a isothermal profile
of temperature $T$.

The total density follows from Eq.(\ref{thermo}). Including the reduction of
the pressure gradient by the turbulent magnetic field with
\begin{equation}
\frac{dp}{dr}=\xi\frac{d}{dr}(\rho_{gas}kT/m )
\end{equation}
we get
\begin{equation}
M(r)=-\frac{\xi kT_0}{Gm}r^2 (1+(r/r_c)^2)^\alpha
\frac{d}{dr}\frac{1}{(1+(r/r_c)^2)^{\alpha(1-\gamma)}}
\end{equation}
or
\begin{equation}
\label{mass}
M(r)=\frac{2\alpha(1-\gamma)\xi kT_0r^3} {Gm(1+(r/r_c)^2)^{1-\alpha\gamma}}
\end{equation}
To fulfil the condition that for large $r$ the gas mass is the dominant
component of the total mass, the exponent of Eq.(\ref{mass}) and
Eq.(\ref{gasmass}) must be the same. That means: $1-\alpha\gamma = \alpha$.
For $\gamma=0.1$ to 0.2 the resulting values are $\alpha=0.91$ to 0.83 (or
$\beta=0.61$ to 0.56). This is exactly the range of values, which is
confirmed by observations.

Compared to equilibrium of an isothermal ideal gas thus for the central
region the total mass is reduced by a factor of $\xi T_0/T$. But now we must
put the question, if this mass is sufficient to keep the galaxies with their
velocities in the range 1000 km/s bound by the gravitational potential of the
cluster plasma. The potential distribution can immediately found by
integration from the equilibrium relation Eq.(\ref{thermo}).
\begin{equation}
\frac{d\Phi}{dr}=-\frac{1}{\rho} \frac{dp}{dr}.
\end{equation}
Introducing the corresponding '$\beta$-profiles' of $\rho$ and $p$ with the
abbreviation $x=r/r_c$ we get
\begin{equation}
\frac{d\Phi}{dx}=\frac{\xi kT_0}{m} \frac{2\alpha(1-\gamma)x}{(1+x^2)^{1-\alpha\gamma}},
\end{equation}
and from this by integration
\begin{equation}
\Phi-\Phi_0= \frac{\xi kT_0}{m}\frac{\alpha(1-\gamma)}{\alpha\gamma}
\left[(1+x^2)^{\alpha\gamma}-1\right].
\end{equation}
or
\begin{equation}
\Phi-\Phi_0=\frac{\xi(1-\gamma)}{\gamma} \frac{ k(T-T_0)}{m}.
\end{equation}
With $\xi=1/3$ and $\gamma=0.1$ we find for a galaxy with an initial radial
velocity $\sigma_r=1000 \rm{km/s}$ that the potential difference, necessary
to slow it down to zero, is
\begin{equation}
\label{flucht}
k(T-T_0)=\frac{m}{3}\frac{\sigma_r^2}{2},
\end{equation}
that means a difference of about 1 keV. On the basis of the empirically
confirmed relation between velocity dispersion and mean gas temperature,
$k\bar{T}=m\sigma_r^2/2$, the condition (\ref{flucht}) is fulfilled for
$\gamma=0.1$ at $r=5.3r_c$, for $\gamma=0.2$ at $r=3.5r_c$. This is just the
region, where the observed density of galaxies in a cluster drops
considerably.

\newpage
\chapter{Idea of a symmetric model of elementary particles}
Under the proposition that spin of elementary particles is not a property of
the constituents, but results from the type of coupling between them, the
complete system of elementary particles can be described by a symmetric
combination of subelementary fields (quarks) and their antifields. To
classify all elementary particles the following rules are sufficient:
\begin{enumerate}
\item
Quarks have a quantum of electric charge or a quantum of strangeness or both.
\item
For every quark there exists an antiquark with the opposite quantum numbers.
\item
All particles consist of equal numbers of quarks and antiquarks.
\item
No two quark can occupy the same quantum state.
\item
Quark couplings can be of "fermionic' or 'bosonic' type.
\end{enumerate}
Fermionic coupling means that the constituent fields are superimposed at the
same location with the resulting spin quantum number 1/2. Bosonic coupling
means that the constituent fields form a spatially extended system, which can
attain different integer rotational quantum numbers.

The conditions described above allow three different quarks, which in the
following are denoted as a, b, and c. They are defined by the quantum numbers
of charge and strangeness:
\begin{equation}
a=\left(\begin{array}{c}1\\-1 \end{array}\right) \quad
b=\left(\begin{array}{c}0\\-1\end{array}\right) \quad
c=\left(\begin{array}{c}1\\0 \end{array} \right)
\end{equation}
As all elementary particles contain equal numbers of quarks and antiquarks,
both systems of quantum numbers are not fixed, but may be varied by an
additional constant. Subtraction of 2/3 from the charge quantum numbers and
adding 1 to the strangeness quantum numbers leads (apart from spin) to the
triplet of antiquarks $\bar{u}$, $\bar{d}$, and$\bar{s}$ of conventional
theory. The negative sign of the strangeness quantum number has been
introduced to remain in agreement with conventional notation.

From combination of two constituents fermionic coupling leads to leptons,
bosonic coupling to mesons. \\\\

\begin {tabular}{ccccc}
Quark-configuration & Charge & Strangeness & Lepton & Meson(Spin 0)\\

$ a \bar{a} $ &0 &0 &$  \nu_{\mu}$ &$ \pi^0  $\\
$ a \bar{b} $ &1 &0 &$ \mu^+$      &$\pi^+ $\\
$ a \bar{c} $ &0 &-1&$ \nu _ e $   &$ \bar{K} ^0 ,\bar{\kappa}$ \\
$ b \bar{a} $ &-1&0 &$\mu^-  $     &$ \pi^-$\\
$ b \bar{b} $ &0 &0 &$ \nu _{\mu}$   &$ \eta $\\
$ b \bar{c} $ &-1&-1&$ e^-  $        &$ K^- $\\
$ c \bar{a} $ &0 &1 & $\bar{\nu}_e$  &$ K^0 , \kappa $\\
$ c \bar{b} $ &1 &1 & $e^+  $        &$ K^+ $\\
$ c \bar{c} $ &0 &0 & $\nu_{\mu} $   &$  \eta '$ \\
\end{tabular}
\\\\
The meson denoted as $ \kappa $ in the table, a hypothetical meson of very
low mass, should exist as a bosonic analog to the electron neutrino. The
existence of such a particle would explain the observed violation of parity
and strangeness conservation in weak interactions. $ K^0 $ mesons can be
regarded as an excited state of the $\kappa$ meson.

The baryon spectrum results from the combination of two quarks and two
antiquarks with the two quarks forming a fermionically bound 'nucleus',
surrounded by a bosonically bound 'shell' of two antiquarks. For the
fermionically bound 'nucleus' the exclusive condition requires that the two
quarks have to be different. Thus only the three combinations (ab), (ac), and
(bc) are possible. For the antiquarks of the shell the exclusion principle is
important only, if both are in the same quantum state. Thus exactly the same
combinations of quantum states are possible, which are found in the
conventional quark model for the ground state octet and singlet. The excited
states form complete decuplets. The ground state multiplet consists of the
following particles: \\\\
\begin{tabular}{cccc}
Quark configuration & Charge & Strangeness & Baryon\\
 $ (ab) \bar{a}  \bar{b}$  & 0 & 0  &$ \Lambda  $\\
 $ (ab) \bar{a}  \bar{c}$ & -1&-1 & $\Xi^-$ \\
 $ (ab) \bar{b}  \bar{c}$ & 0&-1 & $\Xi^0 $ \\
 $ (ac) \bar{a}  \bar{b}$ & 1&1 & p \\
 $ (ac) \bar{a}  \bar{c}$ & 0&0 & $\Sigma^0$ \\
 $ (ac) \bar{b}  \bar{c}$ & 1&0 & $\Sigma^+$\\
 $ (bc) \bar{a}  \bar{b}$ & 0&1 & n   \\
 $ (bc) \bar{a}  \bar{c}$ & -1&0 & $\Sigma^- $\\
 $ (bc)  \bar{b} \bar{c}$ & 0&0 & $\Sigma^0$\\
 \end{tabular}
\\\\
The strangeness numbers in the table differ from conventional notation by 1,
they correspond to the quantum number normally denoted as 'hypercharge'.

Excited states are possible with three additional combinations of quantum
numbers, which result from the configurations $ (ab) \bar{c} \bar {c}' $, $
(ac)\bar{b} \bar{b}' $ and $ (bc) \bar{a} \bar {a}' $ . These configurations
have the quantum numbers (-1,-2), (2,-1) und (-1,-1) and correspond to
$\Omega^- $ , $\Delta^{++} $, and $\Delta^- $.

As an additional combination of four constituents there exists the bosonic
bond of a pair of two fermionically bound quarks with a pair of two
fermionically bound antiquarks. Such particles should exhibit similar
properties as mesons, but with the difference that here the two bosonically
bound building stones are real fermions. Due to the double number of
constituents the mass of these particles should be much higher than that of
normal mesons. To the family of these 'hypermesons' we can assign all the
mesons, to which properties as 'charm ' and 'beauty' are attributed by
conventional theory, as e.g. the mesons $ J/\psi $ und Y.

W and Z particles can be interpreted as fermionic bonds of two quarks and two
antiquarks.
\end{appendix}
\end {document}